\documentclass[10pt,aps,pra,twocolumn,nofootinbib]{revtex4-1}   % style for Physical Review B and AJP are similar

\usepackage[dvips]{epsfig}
\usepackage{amsmath}    % need for subequations
\usepackage{amsfonts}  %note how statements can be commented out
\usepackage{amssymb}
\usepackage{graphicx}   % for figures
\usepackage{mathrsfs}
\usepackage{ulem}
\usepackage{soul}
\usepackage[usenames,dvipsnames]{color}

\usepackage[colorlinks=true,
			urlcolor=blue,
			citecolor=blue,
            linkcolor=blue]{hyperref}   
\usepackage{colortbl}

\definecolor{gray(x11gray)}{rgb}{0.75, 0.75, 0.75}

\definecolor{darksienna}{rgb}{0.24, 0.08, 0.08}

\definecolor{darkblue}{rgb}{0.0, 0.13, 0.35}

\newcommand{\ket}[1]{| #1 \rangle}

\newcommand{\ignore}[1]{}

\newcommand{\be}{\begin{equation}}
\newcommand{\ee}{\end{equation}}
\newcommand{\ba}{\begin{eqnarray}}
\newcommand{\ea}{\end{eqnarray}}

\AtBeginDocument{%
    \hypersetup{
      urlcolor     = blue, %Color for external hyperlinks
      filecolor   = red,
      linkcolor    = blue, %Color of internal links
      citecolor   = blue, %Color of citations
    }
} % end  \AtBeginDocument

\begin{document}

\title{Phase sensitivity of a Mach-Zehnder interferometer with single-intensity and difference-intensity detection}

\author{Stefan~Ataman}
\affiliation{Extreme Light Infrastructure - Nuclear Physics (ELI-NP)\\
30 Reactorului Street, 077125 Bucharest-M\u{a}gurele, Romania}
\email{stefan.ataman@eli-np.ro}

\author{Anca Preda}
\affiliation{Faculty of Physics, University of Bucharest, 077125 Bucharest-M\u{a}gurele, Romania}
\email{anca.preda10@gmail.com}

\author{Radu Ionicioiu}
\affiliation{Horia Hulubei National Institute of Physics and Nuclear Engineering, 077125 Bucharest--M\u agurele, Romania}
\email{r.ionicioiu@theory.nipne.ro}

\date{\today}

% ---------------------------------------------------------
% ----------------------- ABSTRACT ------------------------
% ---------------------------------------------------------

\begin{abstract}
Interferometry is a widely-used technique for precision measurements in both classical and quantum contexts. One way to increase the precision of phase measurements, for example in a Mach-Zehnder interferometer (MZI), is to use high-intensity lasers. In this paper we study the phase sensitivity of a MZI in two detection setups (difference intensity detection and single-mode intensity detection) and for three input scenarios (coherent, double coherent and coherent plus squeezed vacuum). For the coherent and double coherent input, both detection setups can reach the quantum Cram\'er-Rao bound, although at different values of the optimal phase shift. The double coherent input scenario has the unique advantage of changing the optimal phase shift by varying the input power ratio. 
\end{abstract}

\maketitle

% ---------------------------------------------------------
% ------------------- INTRODUCTION ------------------------
% ---------------------------------------------------------
\section{Introduction}
\label{sec:introduction}

Precision measurements are one of key elements in both science and technology. Indeed, many important discoveries have been made due to the improvement of measurement techniques. More sensitive instruments, like microscopes and telescopes, were paramount in discovering new phenomena and in verifying or falsifying theoretical predictions. Thus, improving the measurement sensitivity is a crucial factor driving the advancement of science and technology alike.

A very sensitive, hence widely used measurement technique is interferometry, with the Mach-Zehnder interferometer (MZI) as a standard tool. Thus, understanding, controlling and improving the limits of phase sensitivity of an MZI is an active field of research, both theoretically and experimentally \cite{Bar03,Cav81,Aba11,Dem15,Gao14}. 

Classically, the sensitivity $\Delta\varphi$ of a measurement is bounded by the {\em standard quantum limit} (SQL), also known as the {\em shot-noise limit} \cite{GerryKnight, MandelWolf, Dem15}. This is given by $\Delta\varphi_{SQL}\sim1/\sqrt{\langle{N}\rangle}$, where $\langle{N}\rangle$ is the average number of photons used to probe the system.

It was soon realized that squeezed states of light \cite{Yue76,Yur85,Aga12} can improve the phase sensitivity of an interferometer \cite{Hol93,Par95}. Indeed, this technique has been tested and will be used at the LIGO detector in the future \cite{Aba11,LIGO13}. In a seminal article Caves \cite{Cav81} has showed that squeezed light can improve the phase sensitivity of an interferometer below the shot-noise limit. Experimental demonstration with a MZI \cite{Xia87} soon followed, proving the usability of the concept in practical measurements. Over the next decades both theoretical and experimental studies have showed how to improve the sensitivity of a MZI fed by both a coherent and a squeezed vacuum input \cite{Bre98,Vah10,Vah16,Wak14}.

In a quantum context, however, the phase sensitivity is bounded by the {\em Heisenberg limit} \cite{Ou96,Hol93,Dem15,Gio04,Pez09}, ${\Delta\varphi_{HL}\sim1/\langle{N}\rangle}$, and this limit is fundamental \cite{Gio12}. The so-called NOON states \cite{Hol93,Ou96,Pez09} saturate this limit, while separable states obey the SQL \cite{Pez09}.

The Heisenberg limit can be achieved in a MZI by injecting a coherent state in one port and squeezed vacuum into the other \cite{Pez08}, if roughly half of the input power goes into squeezing. This result was confirmed by Lang and Caves \cite{Lan13, Lan14} who, moreover, showed the input state to be optimal for the class of coherent $\otimes$ squeezed vacuum type of states.

Other scenarios considering active SU(1,1) type interferometers were studied in \cite{Spa15, Spa16}. The authors showed a Heisenberg sensitivity limit achievable in a MZI with squeezed coherent light in both inputs, if the squeezing power is roughly $1/3$ of the total power.

The phase sensitivity of a MZI is not constant \cite{Dem15}. For a small phase variation measurement, one can assume that the interferometer is pre-configured at a convenient point, where the sensitivity is maximal. In order to extend the (rather limited) range of values where each detection scheme approaches the Cram\'er-Rao bound, we can use a Bayesian approach and photon-number resolving detectors. The Cram\'er-Rao bound can be reached with this technique for any value of $\varphi$, as shown by Pezz\'e and Smerzi \cite{Pez07}. Moreover, this can be also achieved for the coherent $\otimes$ squeezed vacuum input \cite{Pez08}.

There are several detection methods used to measure the output of a MZI \cite{Gar17}, however in this paper we shall focus only on two. In the {\em difference intensity detection} scheme, as the name suggests, we have two detectors (one for each output of the MZI) and we measure the difference of the two photo-currents. In the {\em single-mode intensity detection} scheme we measure only one photo-count of the two. For low-power setups the difference intensity detection scheme is experimentally preferred. Here we show that for high input power, the single-mode detection scheme is superior to the difference intensity detection scheme.

We also consider the double coherent input case in this paper. To our best knowledge, this scenario was only discussed by Shin et al. \cite{Shi99}. Moreover, we show that this scenario can have a practical interest under certain circumstances.

Although Heisenberg limited metrology has been a constant theoretical and experimental challenge, this favourable scenario happens for NOON states \cite{Hol93}, where the current record in the number of photons remains very low \cite{Nag07,Afe10} or at extremely low laser powers coupled with the highest squeezing factors achievable today. 

In this paper we are not interested in pursuing the Heisenberg limit at all costs. Instead we focus on scenarios where the squeezing factor is a limited resource, but the intensity of the coherent source is not constrained \cite{Aba11,Ata18b}. This setup is better suited to present-day experiments.

The paper is structured as follows. In Section \ref{sec:MZI_setup_and_sensitivities} we introduce our parameter estimation method, experimental setup, field operator transformations and output operator calculations. We also review the Cram\'er-Rao bound and the Fisher information approach. In Section \ref{sec:coherent_input} we consider a coherent $\otimes$ vacuum input scenario and evaluate its phase sensitivity, comparing both output detection scenarios with the quantum Cram\'er-Rao bound. In Sections \ref{sec:dual_coh_input} and \ref{sec:coherent_plus_squeezed_vacuum} we consider a coherent $\otimes$ coherent and, respectively, coherent $\otimes$ squeezed vacuum input scenarios. We evaluate their respective phase sensitivities, compare the output detection scenarios and assess them in respect with the quantum Cram\'er-Rao bound. All three scenarios are thoroughly discussed and finally, conclusions are drawn in Section \ref{sec:conclusions}.

% ---------------------------------------------------------
% ------------------ MZI SETUP and SENSITIVITIES ----------
% ---------------------------------------------------------
\section{MZI setup: detection sensitivities}
\label{sec:MZI_setup_and_sensitivities}

% ---------------------------------------------------------
% -------------------- PARAMETER ESTIMATION ---------------
% ---------------------------------------------------------
\subsection{Parameter estimation: a short introduction}
\label{subsec:theo_introduction}
\begin{figure}%[h!]
\centering
\includegraphics[scale=0.7]{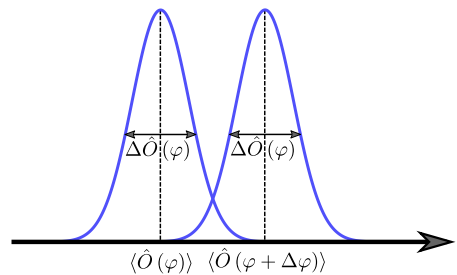}
\caption{\label{fig:average_STD_phi} The physical intuition behind equation \eqref{eq:delta_O_operators_greater_variance_O}. The sensitivity $\Delta\varphi$ depends on both the displacement of the average $\langle\hat{O}\rangle$ (due to a change of the parameter $\varphi$) and the standard deviation $\Delta\hat{O}$. Here we implicitely assume $\Delta\hat{O}(\varphi)= \Delta\hat{O}(\varphi+ \Delta \varphi)$.}
\end{figure}

We now briefly overview the problem of parameter estimation in quantum mechanics. An experimentally accessible Hermitian operator $\hat{O}$ depends on the parameter $\varphi$ -- in our case this is the phase shift in a Mach-Zehnder interferometer; by itself $\varphi$ may or may not be an observable. The average of the operator is
\begin{equation}
\langle\hat{O}\left(\varphi\right)\rangle=\langle\psi\vert\hat{O}\left(\varphi\right)\vert\psi\rangle
\end{equation}
where $\vert\psi\rangle$ is the wave-function of the system. A small variation $\delta\varphi$ of the parameter $\varphi$ induces a change
\begin{equation}
\langle\hat{O}\left(\varphi+\delta\varphi\right)\rangle\approx\langle\hat{O}\left(\varphi\right)\rangle+\frac{\partial\langle\hat{O}\left(\varphi\right)\rangle}{\partial\varphi}\delta\varphi
\end{equation}
The difference $\langle\hat{O}\left(\varphi+\delta\varphi\right)\rangle-\langle\hat{O}\left(\varphi\right)\rangle$ is experimentally detectable if
\begin{equation}
\label{eq:delta_O_operators_greater_variance_O}
\langle\hat{O}\left(\varphi+\delta\varphi\right)\rangle-\langle\hat{O}\left(\varphi\right)\rangle \geq \Delta\hat{O}(\varphi)
\end{equation}
where $\Delta\hat{O}(\varphi):= [\langle\hat{O}^2\rangle- \langle\hat{O}\rangle^2]^{1/2}$ is the standard deviation of $\hat{O}$. One can intuitively understand this condition graphically, see Fig.~\ref{fig:average_STD_phi}. The value of $\delta\varphi$ that saturates the inequality \eqref{eq:delta_O_operators_greater_variance_O} is called {\em sensitivity} and is denoted by $\Delta\varphi$:
\begin{equation}
\label{eq:Delta_varphi_DEFINITION}
\Delta\varphi=\frac{\Delta\hat{O}}{\big\vert\frac{\partial}{\partial\varphi}\langle\hat{O}\rangle\big\vert}
\end{equation}
This equation will be pivotal in the following sections.

% ---------------------------------------------------------
% ------------------------ FIGURE --- FIGURE --- FIGURE ---
% ---------------------------------------------------------

% ---------------------------------------------------------
% ------------ FIELD OPERATOR TRANSFORMATIONS -------------
% ---------------------------------------------------------
\subsection{Transformations of the field operators}
\label{subsec:MZI_field_operator_transf}
\begin{figure}
\centering
\includegraphics[scale=0.5]{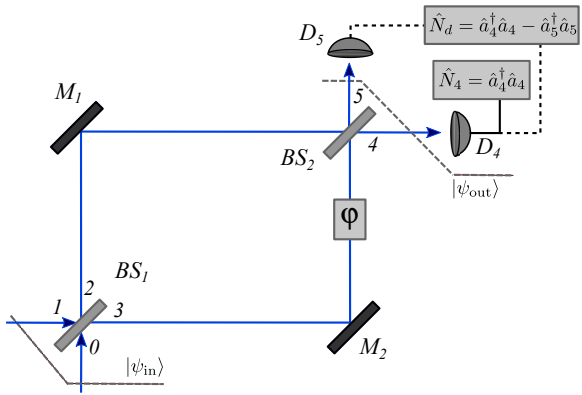}
\caption{\label{fig:MZI_2D_single_diff} The two detection schemes for the Mach-Zehnder interferometer we analyse here. The input state $\vert\psi_{in}\rangle$ is unitarily transformed to the output $\vert\psi_{out}\rangle$. The parameter we want to estimate via a suitable observable is the phase difference $\varphi$ between the two arms of the MZI.}
\end{figure}

Consider a Mach-Zehnder interferometer composed of two mirrors $M_{1,2}$ and two balanced beam splitters $BS_{1,2}$; the transmission (reflection) coefficient of $BS_{1,2}$ is $T=1/\sqrt{2}$ ($R=i/\sqrt{2}$), see Fig.~\ref{fig:MZI_2D_single_diff}. We denote the two input (output) ports by $0$ and $1$ ($4$ and $5$).

The transformation of the field operators between the input and the output of the MZI is
\begin{equation}
\label{eq:field_op_transf_MZI}
\left\{
\begin{array}{l}
\hat{a}_4^\dagger=-\sin\left(\frac{\varphi}{2}\right)\hat{a}_0^\dagger+\cos\left(\frac{\varphi}{2}\right)\hat{a}_1^\dagger\\
\hat{a}_5^\dagger=\mbox{ }\cos\left(\frac{\varphi}{2}\right)\hat{a}_0^\dagger+\sin\left(\frac{\varphi}{2}\right)\hat{a}_1^\dagger
\end{array}
\right.
\end{equation}
where we ignored global phases. We assume the output ports $4$ and $5$ are connected to ideal detectors.

Usually the input state $\vert\psi_{in}\rangle$ is given and we calculate either the output photo-currents or the difference between the output photo-currents.

In the following we denote by $\varphi$ the total phase shift inside the interferometer. The total phase has two parts: (i) the unknown (e.g. sensor-generated) phase shift $\varphi_s$, which is the quantity we want to measure, and (ii) the experimentally-controllable part $\varphi_{exp}$:
\be
\varphi= \varphi_s + \varphi_{exp}
\ee
We assume that $\vert\varphi_s\vert \ll |\varphi|$ so that in order to have the best performance, the experimenter must adjust $\varphi_{exp}$ as close as possible to the optimal phase shift, $\varphi^{\mathrm{opt}}$.

% ---------------------------------------------------------
% ----------------------------------- FIGURE --- FIGURE ---
% ---------------------------------------------------------

% ---------------------------------------------------------
% ------------- OUTPUT OBSERVABLES --- GENERAL ------------
\subsection{Output observables}
Each detection scheme has associated an observable characterising the measurement setup. We will discuss two measurement strategies: (i) difference intensity detection and (ii) single-mode intensity detection scheme.

For Mach-Zehnder interferometers, a well-known approach of calculating the phase sensitivity is Schwinger's scheme based on angular momentum operators \cite{Yur86,Dem15}. Although this method gives faster results for a difference intensity detector setup, it is not well-suited for the single-mode intensity detection scheme we investigate here. Alternatively one can use a Wigner-function based method \cite{Gar17}. In this paper we use a ``brute-force'' calculation based on the field operator transformations \eqref{eq:field_op_transf_MZI}.

% ---------------------------------------------------------
\subsubsection{Difference intensity detection scheme}
In the first detection strategy we calculate the difference between the output photo-currents (i.e., detectors $D_4$ and $D_5$, see Fig.~\ref{fig:MZI_2D_single_diff}). Thus, the observable conveying information about the phase $\varphi$ is
\begin{equation}
\label{eq:N_d_operator_DEFINITION}
\hat{N}_d\left(\varphi\right)=\hat{a}_4^\dagger\hat{a}_4-\hat{a}_5^\dagger\hat{a}_5
\end{equation}
Using the field operator transformations eqs.~\eqref{eq:field_op_transf_MZI} we have
\begin{eqnarray}
\label{eq:Nd_average}
\langle\hat{N}_d\rangle=\cos\varphi\left(\langle\hat{a}_1^\dagger\hat{a}_1\rangle-\langle\hat{a}_0^\dagger\hat{a}_0\rangle\right)
%\nonumber
%\\
-\sin\varphi\left(\langle\hat{a}_0\hat{a}_1^\dagger\rangle
+\langle\hat{a}_0^\dagger\hat{a}_1\rangle\right)
\end{eqnarray}
where the expectation values are calculated w.r.t.~the input state $\ket{\psi_{in}}$. To estimate the phase sensitivity in eq.~\eqref{eq:Delta_varphi_DEFINITION} we need the absolute value of the derivative
\begin{eqnarray}
\label{eq:del_Nd_average_over_del_phi}
\bigg\vert\frac{\partial\langle\hat{N}_d\rangle}{\partial\varphi}\bigg\vert
=\vert\sin\varphi(\langle\hat{a}_0^\dagger\hat{a}_0\rangle-\langle\hat{a}_1^\dagger\hat{a}_1\rangle)
%\nonumber\\
-\cos\varphi(\langle\hat{a}_0\hat{a}_1^\dagger\rangle
+\langle\hat{a}_0^\dagger\hat{a}_1\rangle)\vert
\end{eqnarray}
In the following sections we will calculate this for various input states. The standard deviation ${\Delta\hat{N}_d= [\langle\hat{N}_d^2\rangle-\langle\hat{N}_d\rangle^2]^{1/2}}$ follows from eq.~\eqref{eq:Nd_average} and Appendix \ref{sec:app:variance_calculation}.

% ---------------------------------------------------------
% -------------------- SINGLE DETECTOR --------------------
% ---------------------------------------------------------
\subsubsection{Single-mode intensity detection scheme}
We now consider the single-mode intensity detection scheme, i.e., we have only one detector coupled at the output port $4$, see Fig.~\ref{fig:MZI_2D_single_diff}. Thus the operator of interest is $\hat{N}_4=\hat{a}_4^\dagger\hat{a}_4$. From eq.~\eqref{eq:field_op_transf_MZI} we have
\begin{eqnarray}
\label{eq:N_4_average_GENERAL}
\langle\hat{N}_4\rangle=\sin^2\left(\frac{\varphi}{2}\right)\langle\hat{a}_0^\dagger\hat{a}_0\rangle+\cos^2\left(\frac{\varphi}{2}\right)\langle\hat{a}_1^\dagger\hat{a}_1\rangle
\nonumber\\
-\frac{\sin\varphi}{2}\langle\hat{a}_0\hat{a}_1^\dagger\rangle
-\frac{\sin\varphi}{2}\langle\hat{a}_0^\dagger\hat{a}_1\rangle
\end{eqnarray}
and the absolute value of its derivative w.r.t. $\varphi$ is
\begin{eqnarray}
\label{eq:N4_derivative_average}
\bigg\vert\frac{\partial\langle\hat{N}_4\rangle}{\partial\varphi}\bigg\vert
=\frac{1}{2}\bigg\vert\sin\varphi\left(\langle\hat{a}_0^\dagger\hat{a}_0\rangle-\langle\hat{a}_1^\dagger\hat{a}_1\rangle\right)
\nonumber\\
-\cos\varphi\left(\langle\hat{a}_0\hat{a}_1^\dagger\rangle
+\langle\hat{a}_0^\dagger\hat{a}_1\rangle\right)\bigg\vert
\end{eqnarray}
As before, the standard deviation $\Delta\hat{N}_4$ follows from eq.~\eqref{eq:N_4_average_GENERAL} and Appendix \ref{sec:app:variance_calculation}.

% ---------------------------------------------------------
% --------- ESTIMATION via FISHER INFORMATION -------------
% ---------------------------------------------------------
\subsection{Parameter estimation via Fisher information}
\label{subsec:theo_Fisher}
\begin{figure}%[h!]
\centering
\includegraphics[scale=0.6]{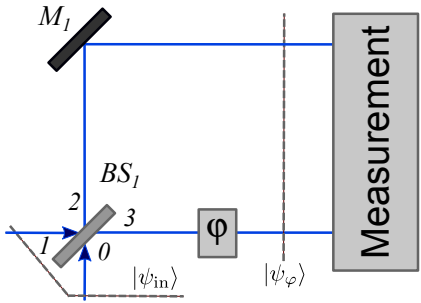}
\caption{\label{fig:MZI_2D_Fisher_info} In the Fisher information approach the Mach-Zehnder interferometer is considered up to the phase shift operation (in our case $\hat{U}\left(\varphi\right)=e^{-i\varphi\hat{a}_3^\dagger\hat{a}_3}$) and the detection scheme completely disregarded; we have ${\ket{\psi_{\varphi}}=\hat{U}\left(\varphi\right)\hat{U}_{BS}\ket{\psi_{in}}}$, where $\hat{U}_{BS}$ is the unitary corresponding to $BS_1$.}
\end{figure}

The Fisher information is a very elegant approach of finding the best-case solution of parameter estimation \cite{Bra94}. The lower bound for the estimation of a parameter $\varphi$ is given by the Cram\'er-Rao bound (CRB) \cite{Dem15, Spa16, Par09}
\be
\Delta\varphi \geq \frac{1}{\sqrt{F\left(\varphi\right)}}
\ee
where $F\left(\varphi\right)$ is the Fisher information. The Fisher information $F\left(\varphi\right)$ is maximised by the quantum Fisher information (QFI) \cite{Bra94} $F\left(\varphi\right)\leq H\left(\varphi\right)$. This leads to the quantum Cram\'er-Rao bound (QCRB)
\begin{equation}
\label{eq:Quantum_Cramer_Rao_Bound}
\Delta\varphi\geq\frac{1}{\sqrt{H\left(\varphi\right)}}
\end{equation}
Here $H(\varphi)= \textrm{Tr}\left[\hat{\rho}_\varphi\hat{L}^2_\varphi\right]$ and $\hat{\rho}_\varphi= \vert\psi_\varphi\rangle\langle\psi_\varphi\vert$ is the density matrix of our system (see Fig.~\ref{fig:MZI_2D_Fisher_info}); $\hat{L}_\varphi$ is the symmetric logarithmic derivative defined as \cite{Bra94,Dem15,Par09} ${\hat{L}_\varphi\hat{\rho}_\varphi+\hat{\rho}_\varphi\hat{L}_\varphi=2\partial\hat{\rho}_\varphi/\partial\varphi}$. Moreover, if the system is in a pure state the quantum Fisher information is  $H(\varphi)= 4\left(\langle\partial_\varphi\psi_\varphi\vert\partial_\varphi\psi_\varphi\rangle-\vert\langle\partial_\varphi\psi_\varphi\vert\psi_\varphi\rangle\vert^2\right)$, where ${\vert\partial_\varphi\psi_\varphi\rangle=\partial\vert\psi_\varphi\rangle/\partial\varphi}$ \cite{Par09,Spa15,Spa16}.

% ---------------------------------------------------------
% ------------------------ FIGURE --- FIGURE --- FIGURE ---
% ---------------------------------------------------------

Importantly, calculating the Fisher information for a given scenario is not always straightforward, and moreover it can lead to different results \cite{Jar12}. Indeed, an external phase reference is needed w.r.t. which are defined the two phase-shifts, each in one arm of the MZI. For this reason, a two parameter estimation problem involving a Fisher matrix is used \cite{Lan13}, see Appendix \ref{sec:app:Dual_coh_QCRB}. When an external phase reference is not available, one has to pay particular attention on what is actually measurable given the experimental setup \cite{Tak17}.

We stress that in the evaluation of QCRB the detection scheme is disregarded, see Fig.~\ref{fig:MZI_2D_Fisher_info}. The QCRB will always be a theoretical, best case scenario, which overlooks practical implementations of the detection stage.

In the following, for each case discussed in Sections \ref{sec:coherent_input}, \ref{sec:dual_coh_input} and \ref{sec:coherent_plus_squeezed_vacuum} we will compare the practically achievable results with the QCRB from eq.~\eqref{eq:Quantum_Cramer_Rao_Bound}.

% ---------------------------------------------------------
% -------------------- SINGLE COHERENT INPUT --------------
% ---------------------------------------------------------
\section{Single coherent input}
\label{sec:coherent_input}
In this section we consider the input port $1$ in a coherent state $\vert\alpha\rangle$ while input port $0$ is kept ``dark'' (i.e., in the vacuum state). The input state is
\be
\label{eq:psi_in_coherent}
\ket{\psi_{in}}= \ket{\alpha_10_0}= \hat{D}_1 (\alpha) \ket{0}
\ee
where $\hat{D}_1\left(\alpha\right)=e^{\alpha\hat{a}_1^\dagger-\alpha^*\hat{a}_1}$ is the displacement operator \cite{GerryKnight,MandelWolf,Aga12}.

% ----------------- SINGLE COHERENT INPUT -----------------
% ----------------- DIFFERENTIAL DETECTION ----------------
% ---------------------------------------------------------
\subsection{Difference intensity detection scheme}
\label{subsec:coherent_input_diff_det}
The observable we measure is the difference in the photo-currents at the outputs $4$ and $5$, namely the average value of $\hat{N}_d$, eq.~\eqref{eq:N_d_operator_DEFINITION}. For the input state \eqref{eq:psi_in_coherent} we find ${\langle\hat{N}_d\rangle= \cos\varphi \vert\alpha\vert^2}$ and, using equation \eqref{eq:N_d_SQUARED_FINAL}, the output variance is found to be ${\Delta^2\hat{N}_d=\vert\alpha\vert^2}$. Consequently, the phase sensitivity of a Mach-Zehnder interferometer driven by a single coherent source is
\begin{equation}
\label{eq:Delta_varphi_singleCOH_Difference_FINAL}
\Delta\varphi_{\mathrm{diff}}
=\frac{1}{\vert\sin\varphi\vert \vert\alpha\vert}
=\frac{1}{\vert\sin\varphi\vert\sqrt{\langle{N}\rangle}}
\end{equation}
where the average number of photons is $\langle{N}\rangle=\vert\alpha\vert^2$ and this is the well-known shot noise limit or standard quantum limit \cite{Dem15,Pez07}.

% ------------------ SINGLE COHERENT INPUT ----------------
% --------------------- SINGLE DETECTOR -------------------
% ---------------------------------------------------------
\subsection{Single-mode intensity detection scheme}
\label{subsec:SINGLE_COH_single_detector}
In a single-mode intensity detection setup the average of the output observable $\hat{N}_4$ gives
\begin{equation}
\label{eq:N4_average_COHERENT}
\langle\hat{N}_4\rangle=\cos^2\left(\frac{\varphi}{2}\right)\vert\alpha\vert^2
\end{equation}
The variance of $\hat{N}_4$ follows from eqs.~\eqref{eq:N_4_SQUARED_FINAL} and \eqref{eq:N4_average_COHERENT}, giving ${\Delta^2\hat{N}_4=\cos^2\left({\varphi}/{2}\right)\vert\alpha\vert^2}$. Thus, the phase sensitivity in the single-mode intensity detection case is
\begin{equation}
\label{eq:Delta_varphi_N4_single_coherent}
\Delta\varphi_{\mathrm{sing}}
=\frac{1}{\vert\sin\left(\frac{\varphi}{2}\right)\vert \vert\alpha\vert}
=\frac{1}{\vert\sin\left(\frac{\varphi}{2}\right)\vert \sqrt{\langle{N}\rangle}}
\end{equation}

\begin{figure}
	\centering
	\includegraphics[scale=0.45]{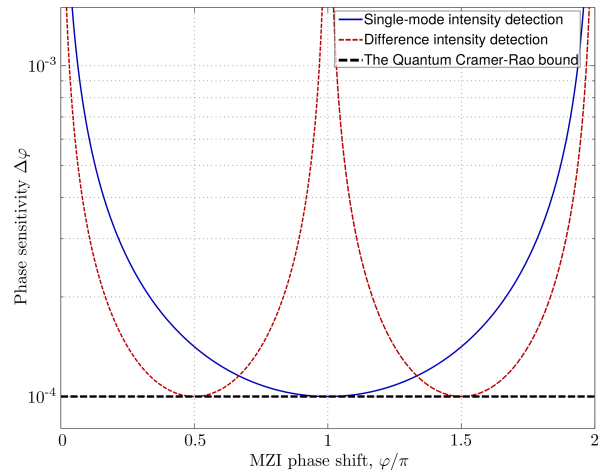}	
	\caption{Phase sensitivity for the single-mode (solid blue line) and difference (dashed red line) intensity detection setups compared to the quantum Cram\'er-Rao bound (thick dashed line) for a single coherent input with $\vert\alpha\vert=10^4$. Both configurations reach the Cram\'er-Rao bound at their respective optimal phase shifts.}
	\label{fig:Plot_Coh_Diff_Single_CRB_var_phase}
\end{figure}

% ------------ COMPARISON with QUANTUM CRAMER-RAO ---------
% ---------------------------------------------------------
\subsection{Discussion: the quantum Cram\'er-Rao bound}
\label{subsec:SINGLE_COH_discussion}

For a single input coherent state, the QCRB in equation \eqref{eq:Quantum_Cramer_Rao_Bound} is \cite{Dem15,Jar12}
\begin{equation}
\label{eq:Delta_varphi_CramerRaoBound}
\Delta\varphi_{\mathrm{QCRB}}
\geq\frac{1}{\vert\alpha\vert}
=\frac{1}{\sqrt{\langle{N}\rangle}}
\end{equation}
Both detection schemes reach this limit, but at different values of the total internal phase shift, as depicted in Fig.~\ref{fig:Plot_Coh_Diff_Single_CRB_var_phase}.

In the differential detection scheme, the optimal sensitivity is reached for $|\sin \varphi|= 1$, i.e., $\varphi_{\mathrm{diff}}^{\mathrm{opt}}= \pi/2+k\pi$, $k\in\mathbb{Z}$, see eq.~\eqref{eq:Delta_varphi_singleCOH_Difference_FINAL}. This implies equal output power at the two outputs (4 and 5). There is no ``dark'' port in the case of the difference intensity detection. This can be a major drawback if one uses a high input power in order to lower the sensitivity.

For single-mode intensity detection, the phase sensitivity \eqref{eq:Delta_varphi_N4_single_coherent} reaches the QCRB at ${\varphi_{\mathrm{sing}}^{\mathrm{opt}}=\pi+2k\pi}$, $k\in\mathbb{Z}$, see Fig.~\ref{fig:Plot_Coh_Diff_Single_CRB_var_phase}. This means that the output $4$ is a ``dark'' port. This is a clear advantage for high input power since we can use extremely sensitive PIN photo-diodes \cite{Bro10}.

% ---------------------------------------------------------
% ----------------- DOUBLE COHERENT LIGHT -----------------
% ---------------------------------------------------------
\section{Double coherent input}
\label{sec:dual_coh_input}
An interesting situation arises if we apply a coherent source in each input port of the interferometer:
\begin{equation}
\label{eq:psi_in_double_coherent}
\vert\psi_{in}\rangle=\vert\alpha_1\beta_0\rangle=\hat{D}_1\left(\alpha\right)\hat{D}_0\left(\beta\right)\vert0\rangle
\end{equation}
where the displacement operator at input port $0$ is ${\hat{D}_0\left(\beta\right)= e^{\beta\hat{a}_0^\dagger-\beta^*\hat{a}_0}}$. Here $\alpha= \vert\alpha\vert e^{i\theta_\alpha}$, $\beta= \vert\beta\vert e^{i\theta_\beta}$ and $\Delta\theta= \theta_\alpha-\theta_\beta$ is the phase difference between the two input lasers.

% --------------- DOUBLE COHERENT LIGHT -------------------
% --------------- DIFFERENTIAL DETECTION ------------------
% ---------------------------------------------------------
\subsection{Differential detection scheme}

Using the input state given in equation \eqref{eq:psi_in_double_coherent}, the average value of the operator $\hat{N}_d$ is
\begin{equation}
\label{eq:Nd_average_doubleCOH}
\langle\hat{N}_d\rangle=\cos\varphi\left(\vert\alpha\vert^2-\vert\beta\vert^2\right)-2\sin\varphi\vert\alpha\beta\vert\cos\Delta\theta
\end{equation}
After a straightforward computation, the variance is
\begin{eqnarray}
\label{eq:Delta_Nd_2_FINAL_dualCOH}
% ---- SIMPLIFY 2
\Delta^2\hat{N}_d^2=\vert\alpha\vert^2+\vert\beta\vert^2= \vert\alpha\vert^2\left(1+\varpi^2\right)
\end{eqnarray}
where $\varpi:= |\beta|/|\alpha|$. The phase sensitivity for a double coherent input is
\begin{equation}
\label{eq:Delta_phi_dualCOH_Diff_det}
\Delta\varphi_\mathrm{diff}= \frac{\sqrt{1+\varpi^2}}
{\vert\alpha\vert\big\vert\sin\varphi\left(1 -\varpi^2\right)+ 2\cos\varphi\varpi\cos\Delta\theta\big\vert}
\end{equation}
We will discuss this result in Section \ref{subsec:Dual_coh_QCRB}.

% -------------------- DOUBLE COHERENT LIGHT --------------
% --------------------- SINGLE DETECTOR -------------------
% ---------------------------------------------------------
\subsection{Single-mode intensity detection scheme}
In the single-mode intensity detection setup, the average of our output observable is
\begin{eqnarray}
\label{eq:N_4_average_dualCOH}
\langle\hat{N}_4\rangle=\vert\alpha\vert^2\left(\sin^2\left(\frac{\varphi}{2}\right)\varpi^2+\cos^2\left(\frac{\varphi}{2}\right)
\right.
\nonumber\\
%\left.
-\sin\varphi\varpi\cos\Delta\theta\Big)
\end{eqnarray}
The variance $\Delta^2\hat{N}_4$ can be computed as before; alternatively, we notice that at the output port $4$ we have a coherent state, therefore the variance is equal to its average value,
\begin{equation}
\Delta^2\hat{N}_4=\langle\hat{N}_4\rangle
\end{equation}
Thus, the phase sensitivity of a Mach-Zehnder with two input coherent sources and a single-mode intensity detection scheme is
\begin{equation}
\label{eq:Delta_phi_dualCOH_Sing_det}
\Delta\varphi_\mathrm{sing}=\frac{\sqrt{\sin^2\left(\frac{\varphi}{2}\right)\varpi^2
+\cos^2\left(\frac{\varphi}{2}\right)
-\sin\varphi\varpi\cos\Delta\theta}}
{\vert\alpha\vert\Big\vert\frac{\sin\varphi}{2}\left(1-\varpi^2\right)
+\cos\varphi\varpi\cos\Delta\theta\Big\vert}
\end{equation}

% ----------------- DOUBLE COHERENT LIGHT -----------------
% ---------- COMPARISON with QUANTUM CRAMER-RAO -----------
% ---------------------------------------------------------
\subsection{Discussion: the quantum Cram\'er-Rao bound}
\label{subsec:Dual_coh_QCRB}

For the double coherent input, the quantum Cram\'er-Rao bound is (see Appendix \ref{sec:app:Dual_coh_QCRB})
\begin{eqnarray}
\label{eq:Delta_varphi_QCRB_Dual_coh}
\Delta\varphi_{\mathrm{QCRB}}
\geq \frac{1}{\vert\alpha\vert\sqrt{1+ \varpi^2-\frac{4\varpi^2}{1+ \varpi^2}\sin^2\Delta\theta}}
\end{eqnarray}
Therefore the best sensitivity is achieved when the two input lasers are in phase, $\Delta\theta=0$, resulting in ${\Delta\varphi_{QCRB}=1/(\vert\alpha\vert\sqrt{1+\varpi^2}})$. 

In the case of differential detection, one can show that an optimum phase shift exists,
\begin{equation}
\label{eq:phi_OPT_diff_det_Dual_coh}
\varphi^{\textrm{opt}}_\mathrm{diff}= \pm\arctan\left(\frac{\vert1-\varpi^2\vert}{2\varpi\vert\cos\Delta\theta\vert}\right)+k\pi
\end{equation}
with $k\in\mathbb{Z}$ and $\varphi_\mathrm{diff}^{\mathrm{opt}}$ brings the sensitivity form equation \eqref{eq:Delta_phi_dualCOH_Diff_det} to the QCRB.

For the single-mode intensity detection scheme, if the two input lasers are in phase (${\Delta\theta=0}$), the optimum phase shift is
\begin{equation}
\label{eq:phi_OPT_sing_det_Dual_coh}
\varphi^{\textrm{opt}}_\mathrm{sing}= \pm2\arctan\left(\frac{1}{\varpi}\right)+2k\pi
\end{equation}
with $k\in\mathbb{Z}$ and substituting this value into equation \eqref{eq:Delta_phi_dualCOH_Sing_det} gives the QCRB from equation \eqref{eq:Delta_varphi_QCRB_Dual_coh}.

For comparison, the sensitivity of homodyne detection with $\Delta\theta=0$ is
\begin{equation}
\Delta\varphi_H\geq\frac{1}{\vert\alpha\sin\frac{\varphi}{2}+\beta\cos\frac{\varphi}{2}\vert}
% NOTE: the phase between alpga and beta should be included
%=\frac{1}{\vert\alpha\vert \vert\sin\frac{\varphi}{2}+\varpi\cos\frac{\varphi}{2}\vert}
\end{equation}

\begin{figure}
	\centering
	\includegraphics[scale=0.45]{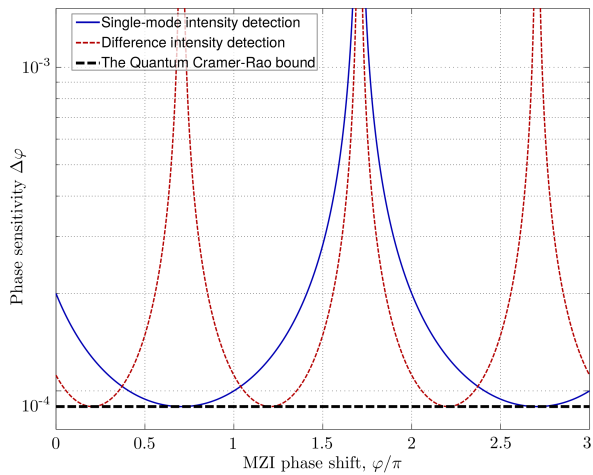}		
	\caption{Phase sensitivity for the single-mode intensity (solid blue line) and difference intensity (dashed red line) detection setups versus the phase shift $\varphi$. Both detection schemes reach the quantum Cram\'er-Rao bound (thick dashed line). We used the parameters $\vert\alpha\vert=10^4$, $\varpi=0.5$ and $\Delta\theta= 0$.}
	\label{fig:D_phi_dual_coh_Sing_Diff_Det_vs_phi}
\end{figure}

The phase sensitivity of a MZI with a double-coherent input is shown in Fig.~\ref{fig:D_phi_dual_coh_Sing_Diff_Det_vs_phi}, for both single-mode and difference intensity detection schemes. As already discussed, we can reach the QCRB in both scenarios.

Compared to the single-coherent input, the double-coherent case has an important advantage: we can tune the value of $\varphi_\mathrm{sing}^{\mathrm{opt}}$ at which the sensitivity reaches the QCRB. Experimentally, this can be achieved by varying the power ratio of the two input coherent sources. This avoids the use of piezos or other mechanical-based methods to induce phase shifts. As a consequence, our proposal reduces mechanical vibrations, noise or mis-alignments.

In the high-power regime this ability is practically useful for a single-mode intensity detection scenario. Indeed, at the optimal phase shift the output 4 is a dark port, i.e., $\langle \hat{N}_4 \left(\varphi_\mathrm{sing}^\mathrm{opt}\right) \rangle\to 0$, which is exactly the desired situation w.r.t. the photo-detectors in the high-power regime.

% ---------------------------------------------------------
% --------------- COHERENT plus SQUEEZED VACUUM -----------
% ---------------------------------------------------------
\section{Coherent plus squeezed vacuum input}
\label{sec:coherent_plus_squeezed_vacuum}
The paradigmatic input state which beats the SQL is the coherent $\otimes$ squeezed vacuum
\begin{equation}
\label{eq:psi_in_coherent_squeezed}
\vert\psi_{in}\rangle=\vert\alpha_1r_0\rangle=D_1\left(\alpha\right)S_0\left(r\right)\vert0\rangle
\end{equation}
The squeezed vacuum state is obtained by applying the squeezing operator $S_0\left(\xi\right)=e^{[(\xi^*)^2\hat{a}_1^2-\xi^2(\hat{a}_1^\dagger)^2]/2}$ \cite{GerryKnight,Yue76,Aga12} with $\xi=re^{i\theta}$. For simplicity, in the following we take ${\theta=0}$, hence $\xi=r\in\mathbb{R}^{+}$. This input state is of considerable practical interest as it was shown to beat the SQL \cite{Cav81,Hol93,Par95,Lan13,Lan14}, a prediction amply confirmed by experiments \cite{Xia87,LIGO13,Aba11,Sch17}.

% ---------------- DIFFERENTIAL DETECTION -----------------
% ---------------------------------------------------------
\subsection{Difference intensity detection scheme}
\label{subsec:CoherentSqueezedVac_differential_det}
With the coherent $\otimes$ squeezed vacuum input \eqref{eq:psi_in_coherent_squeezed} the average of $\hat{N}_d$ in eq.\eqref{eq:N_d_operator_DEFINITION} is 
\be
\label{eq:Nd_average_coherent_squeezed}
\langle\hat{N}_d\rangle=\cos\varphi\left(\vert\alpha\vert^2-\sinh^2r\right)
\ee
The variance $\Delta^2\hat{N}_d$ can be computed using equations \eqref{eq:N_d_operator_DEFINITION} and \eqref{eq:N_d_SQUARED_FINAL} with the input state given in \eqref{eq:psi_in_coherent_squeezed} and yields
\begin{eqnarray}
\label{eq:Delta_Nd_squared_sqz_coh}
\Delta^2\hat{N}_d
=\cos^2\varphi\left(\frac{\sinh^22r}{2}+\vert\alpha\vert^2\right)
\nonumber\\
+\sin^2\varphi\left(\sinh^2r+\vert\alpha\vert^2e^{-r}\right)
\nonumber\\
+\sin^2\varphi\vert\alpha\vert^2\sinh2r\left(1
-\cos\left(2\theta_\alpha\right)\right)
%\qquad
\end{eqnarray}
For the difference intensity detection scheme, the best achievable phase sensitivity of a MZI with coherent $\otimes$ squeezed vacuum is
\begin{widetext}
\begin{eqnarray}
\label{eq:Delta_varphi_CSV_diff_GENERAL}
\Delta\varphi_\mathrm{diff}
=\frac{\sqrt{\left(\vert\alpha\vert^2+\frac{\sinh^22r}{2}\right)\cot^2\varphi
+\sinh^2r+\vert\alpha\vert^2e^{-2r}+\vert\alpha\vert^2\sinh2r\left(1
-\cos\left(2\theta_\alpha\right)\right)}}
{\vert\sinh^2r-\vert\alpha\vert^2\vert}
\end{eqnarray}
The last term in the numerator of equation \eqref{eq:Delta_varphi_CSV_diff_GENERAL} is the input noise enhancement due to the misalignment of the coherent input with respect to the squeezed vacuum (whose phase we considered to be zero, for simplicity). The sensitivity is minimized if the phase of the coherent light is $\theta_\alpha=0$ (hence $\alpha\in\mathbb{R}$):

\begin{eqnarray}
\label{eq:Delta_varphi_Coh_Sqz_Vac_DIFF}
\Delta\varphi_\mathrm{diff}
=\frac{\sqrt{\left(\alpha^2+\frac{\sinh^22r}{2}\right)\cot^2\varphi
+\sinh^2r+\alpha^2e^{-2r}}}
{\vert\alpha^2-\sinh^2r\vert}\quad
\end{eqnarray}
expression that can be found in the literature \cite{Dem15,Pez08}.

%\end{widetext}

% ----------------- SINGLE DETECTOR -----------------------
% ---------------------------------------------------------
\subsection{Single-mode intensity detection scheme}
\label{subsec:CoherentSqueezedVac_single_det}
For the input state \eqref{eq:psi_in_coherent_squeezed} we have
\begin{equation}
\label{eq:N4_average_squeezed_coherent}
\langle\hat{N}_4\rangle=\sin^2\left(\frac{\varphi}{2}\right)\sinh^2r+\cos^2\left(\frac{\varphi}{2}\right)\vert\alpha\vert^2
\end{equation}
and the variance is
%\begin{widetext}
\begin{eqnarray}
\label{eq:Delta_N4_FINAL_sqz_coh}
\Delta^2\hat{N}_4
=\sin^4\left(\frac{\varphi}{2}\right)\frac{\sinh^22r}{2}
+\sin^2\left(\frac{\varphi}{2}\right)\cos^2\left(\frac{\varphi}{2}\right)\sinh^2r+\cos^4\left(\frac{\varphi}{2}\right)\vert\alpha\vert^2
% -----%----- SECOND
\nonumber\\
+\sin^2\left(\frac{\varphi}{2}\right)\cos^2\left(\frac{\varphi}{2}\right)\vert\alpha\vert^2e^{-2r}
+\frac{\sin^2\varphi}{4}\sinh{2r}\vert\alpha\vert^2\left(1-\cos2\theta_\alpha\right)
\end{eqnarray}
In the single-mode intensity detection setup, the best achievable sensitivity of a MZI fed by coherent $\otimes$ squeezed vacuum is
\begin{eqnarray}
\label{eq:Delta_phi_sqz_coh_SINGLE_DET}
\Delta\varphi_\mathrm{sing}
=\frac{\sqrt{\frac{\tan^2\left(\frac{\varphi}{2}\right)\sinh^22r}{2}
+\sinh^2r
% -----%----- SECOND
+\frac{\vert\alpha\vert^2}{\tan^2\left(\frac{\varphi}{2}\right)}
+\vert\alpha\vert^2e^{-2r}
% -----%----- LAST
+\sinh{2r}\vert\alpha\vert^2\left(1-\cos2\theta_\alpha\right)}}
{\Big\vert \vert\alpha\vert^2-\sinh^2r\Big\vert}
%\nonumber
\end{eqnarray}
\end{widetext}
The last term of the square root is again the contribution of the misalignment of the coherent input from port $1$ with the squeezed vacuum from port $0$. The sensitivity is maximized for $\cos2\theta_{\alpha}=1$, thus $\theta_{\alpha}=0$ and hence $\alpha \in \mathbb{R}$. Therefore, we have now the best achievable sensitivity for the squeezed $\otimes$ coherent input and a single-mode intensity detection scheme \cite{Ata18b}
\begin{eqnarray}
\label{eq:Delta_phi_sqz_coh_SINGLE_DET_theta_zero}
\Delta\varphi_\mathrm{sing}
=\frac{\sqrt{\frac{\tan^2\left(\frac{\varphi}{2}\right)\sinh^22r}{2}
+\sinh^2r
% -----%----- SECOND
+\frac{\alpha^2}{\tan^2\left(\frac{\varphi}{2}\right)}
+\alpha^2 e^{-2r}
}}
{\big\vert\alpha^2- \sinh^2 r\big\vert}\quad\:\:
\end{eqnarray}

% -------------- COMPARISON with QUANTUM CRAMER-RAO -------
% ---------------------------------------------------------
\subsection{Discussion: the quantum Cram\'er-Rao bound}
The quantum Cram\'er-Rao bound for the coherent $\otimes$ squeezed vacuum input is \cite{Pez08,Jar12,Lan13,Lan14}
\begin{equation}
\label{eq:Delta_phi_coherent_squeezed_CRB}
\Delta\varphi_{QCRB}\geq\frac{1}{\sqrt{\vert\alpha\vert^2e^{2r}+\sinh^2r}}
\end{equation}
and is independent of the phase shift $\varphi$ of the MZI, similar to the coherent input case.

\begin{figure}
	\centering
	\includegraphics[scale=0.45]{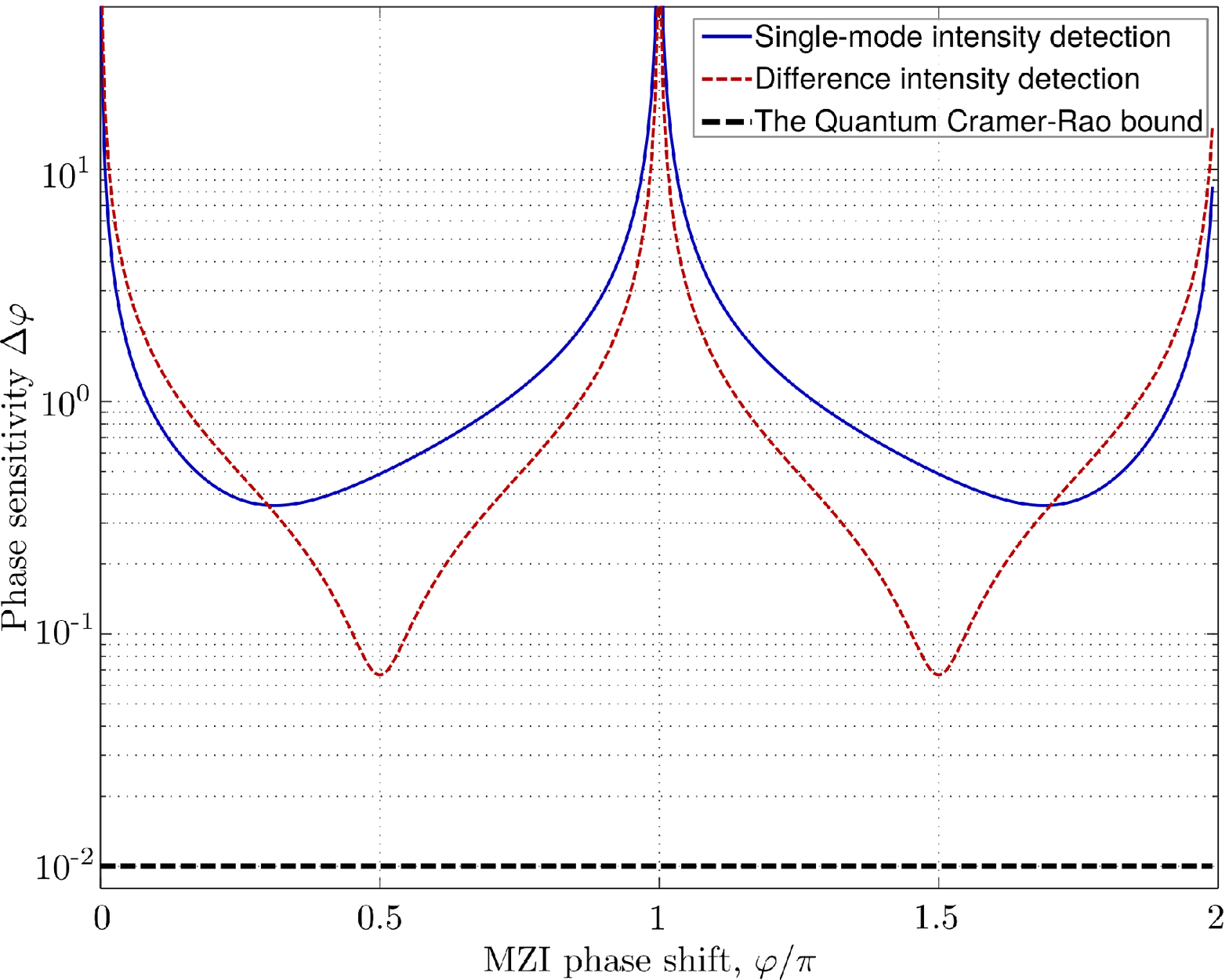}
	\caption{Phase sensitivity for the single-mode (solid blue line) and difference (dashed red line) intensity detection setups compared to the quantum Cram\'er-Rao bound (thick dashed line) versus the phase shift $\varphi$. Here $\vert\alpha\vert=10$ and $r=2.3$.}
	\label{fig:Plot_Coh_SqzVac_Diff_Single_CRB_var_phase_low_alpha}
\end{figure}

For comparison, we briefly mention the sensitivity of the homodyne detection scheme \cite{Gar17,Li14}
\begin{equation}
\label{eq:Delta_phi_coherent_squeezed_Homodyne}
\Delta\varphi_{H}\geq\frac{e^{-r}}{\vert\alpha\vert}
\end{equation}
a result we will use later.

\begin{figure}
	\centering
		\includegraphics[scale=0.45]{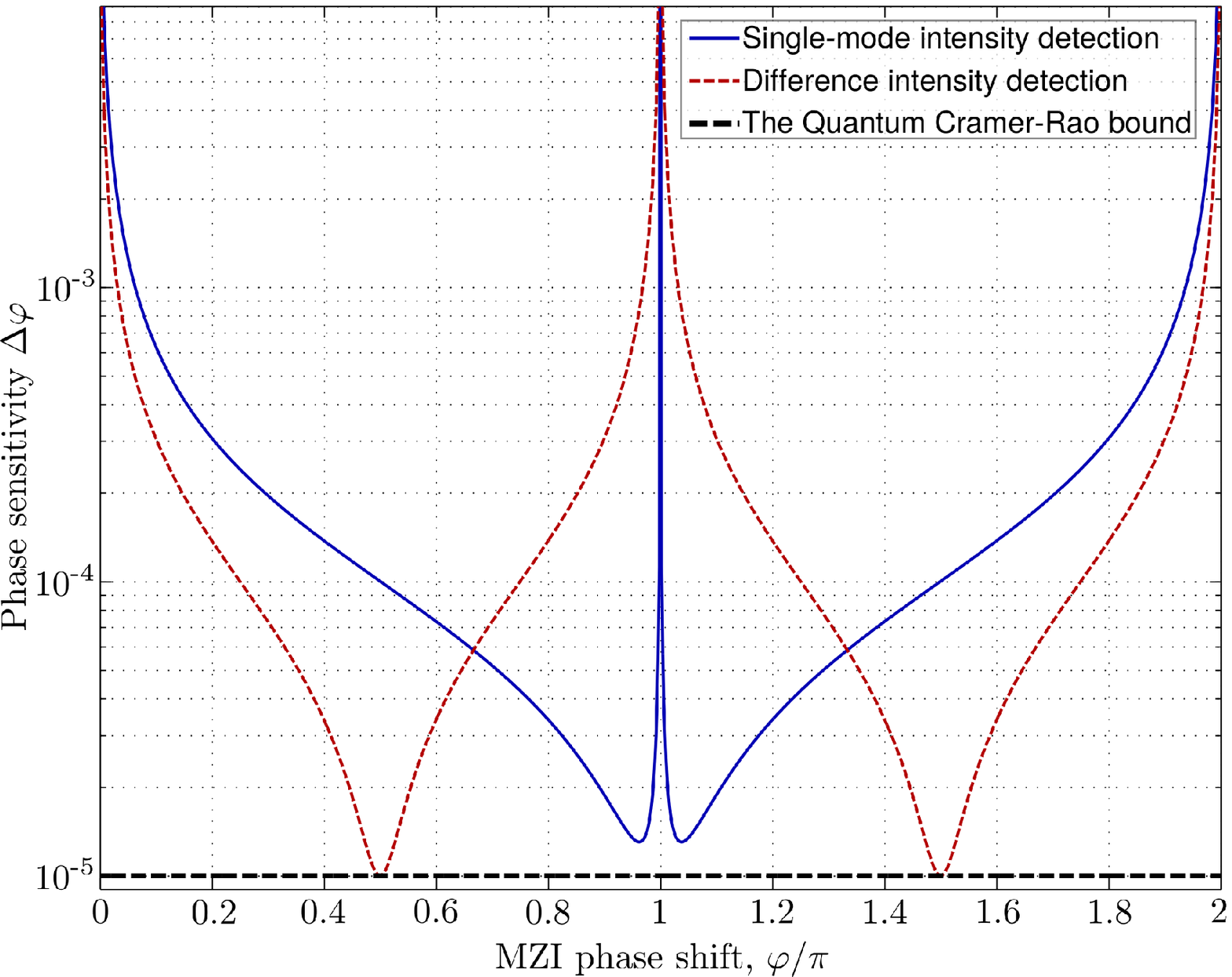}
	\caption{Phase sensitivity for the single-mode (solid blue line) and difference (dashed red line) intensity detection setups compared to the quantum Cram\'er-Rao bound (thick dashed line) versus the phase shift $\varphi$. We use $\vert\alpha\vert=10^4$ and $r=2.3$.}
	\label{fig:Plot_Coh_SqzVac_Diff_Single_CRB_var_phase_high_alpha}
\end{figure}

For the differential detection scheme, the optimal sensitivity in eq.~\eqref{eq:Delta_varphi_Coh_Sqz_Vac_DIFF} is reached for ${\cos\varphi=0}$, i.e., ${\varphi_\mathrm{diff}^{\mathrm{opt}}=\pi/2+k\pi}$, $k\in\mathbb{Z}$ and we find the best achievable sensitivity
\begin{eqnarray}
\Delta\varphi_\mathrm{diff}
=\frac{\sqrt{\sinh^2r+\alpha^2e^{-2r}
}}{\vert\alpha^2-\sinh^2r\vert}
\end{eqnarray}
a result also found in the literature \cite{Pez08,Dem15,Gar17}. 

For single-mode intensity detection, the optimal sensitivity from  equation \eqref{eq:Delta_phi_sqz_coh_SINGLE_DET_theta_zero} is reached when
\begin{equation}
\label{eq:phi_optimal_CSV_single_DET}
\varphi_{\mathrm{sing}}^{\mathrm{opt}}=\pm2\arctan\left(\sqrt{\frac{\sqrt{2}\vert\alpha\vert}{\sinh2r}}\right)+2k\pi
\end{equation}
with $k\in\mathbb{Z}$; substituting this value in equation \eqref{eq:Delta_phi_sqz_coh_SINGLE_DET_theta_zero} gives the best achievable sensitivity in the case of a single-mode intensity scheme, namely
\begin{eqnarray}
\label{eq:Delta_phi_sqz_coh_SINGLE_DET_OPTIMAL}
\Delta\varphi_\mathrm{sing}
=\frac{\sqrt{\sinh^2r+\sqrt{2}\alpha\sinh2r+\alpha^2e^{-2r}}}
{\big\vert\alpha^2-\sinh^2r\big\vert}
\end{eqnarray}
This result is identical to the one reported in reference \cite{Gar17}, equation (10).

In Figs.~\ref{fig:Plot_Coh_SqzVac_Diff_Single_CRB_var_phase_low_alpha} and \ref{fig:Plot_Coh_SqzVac_Diff_Single_CRB_var_phase_high_alpha} we plot the best achievable phase sensitivity in the single-mode and difference intensity detection schemes together with the Cram\'er-Rao bound from equation \eqref{eq:Delta_phi_coherent_squeezed_CRB} for coherent $\otimes$ squeezed vacuum input versus the phase shift of the MZI. One notes that both detection schemes have an optimum, however none reaches the QCRB. (Although in Fig.~\ref{fig:Plot_Coh_SqzVac_Diff_Single_CRB_var_phase_high_alpha} it seems that the red curve corresponding to the difference intensity detection scenario reaches the QCRB, it actually stays above it.) While the optimum working point for the difference intensity detection scheme is constant, in the transition from the low- (Fig.~\ref{fig:Plot_Coh_SqzVac_Diff_Single_CRB_var_phase_low_alpha}) to the high-power regime (Fig.~\ref{fig:Plot_Coh_SqzVac_Diff_Single_CRB_var_phase_high_alpha}) the optimum working point $\varphi^{\mathrm{opt}}$ shifts, see eq.~\eqref{eq:Delta_phi_sqz_coh_SINGLE_DET_OPTIMAL}.

\begin{figure}
	\centering
		\includegraphics[scale=0.45]{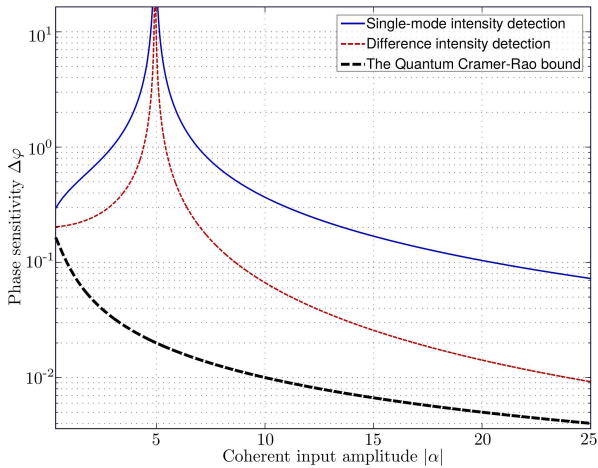}
	\caption{Phase sensitivity for the single-mode  (solid blue line), difference (dashed red line) intensity detection schemes and the quantum Cram\'er-Rao bound (thick dashed line) versus the coherent input amplitude $\vert\alpha\vert$, in the low-intensity regime. We take the squeezing factor $r=2.3$ and consider the optimal phases $\varphi^{\mathrm{opt}}$ for each detection scenario. Both detection techniques are sub-optimal w.r.t. the QCRB, yielding poor performance especially when ${\vert\alpha\vert^2\approx\sinh^2r}$.}
	\label{fig:Plot_Coh_SqzVac_Diff_Single_CRB_var_alpha_low}
\end{figure}

In Fig.~\ref{fig:Plot_Coh_SqzVac_Diff_Single_CRB_var_alpha_low} we show both $\Delta\varphi_\mathrm{diff}$ and $\Delta\varphi_\mathrm{sing}$ in the low $\vert\alpha\vert$ regime. For $\vert\alpha\vert^2\approx\sinh^2{r}$ both detection schemes give poor results while the QCRB reaches the Heisenberg limit $\Delta\varphi_{QCRB}\sim1/\langle{N}\rangle$, where $\langle{N}\rangle=\vert\alpha\vert^2+\sinh^2r$. This behaviour has been explained by Pezz\'e and Smerzi \cite{Pez08} and was attributed to the limited information gained by these phase estimation techniques, notably due to the ignorance of the fluctuation in the number of particles.

% ---------------------------------------------------------
% ------------------------ FIGURE --- FIGURE --- FIGURE ---
% ---------------------------------------------------------

Ideally one would like to enhance the squeezing factor $r$ as much as possible. However, this is experimentally challenging \cite{Vah10,Vah16,Sch17}. The maximum reported squeezing was $15$ dB corresponding to $r\approx 2.3$ \cite{Vah16}. Therefore, in order to remain realistic in the high-intensity regime, wee keep $r$ constant and small compared to the amplitude of the coherent state, implying ${\vert\alpha\vert^2\gg\vert\alpha\vert\gg\sinh^2r}$.  Indeed, for $\vert\alpha\vert\gg\sinh{r}$ both detection schemes equal the sensitivity of the homodyne detection in eq.~\eqref{eq:Delta_phi_coherent_squeezed_Homodyne}. The QCRB in eq.~\eqref{eq:Delta_phi_coherent_squeezed_CRB} can be approximated by $\Delta\varphi\approx{e^{-r}}/{\vert\alpha\vert}$. Thus, using squeezing in port $0$ brings a factor of $e^{-r}$ over the SQL, therefore the coherent $\otimes$ squeezed vacuum technique remains interesting even for large $\vert\alpha\vert$.

\begin{figure}
	\centering
		\includegraphics[scale=0.45]{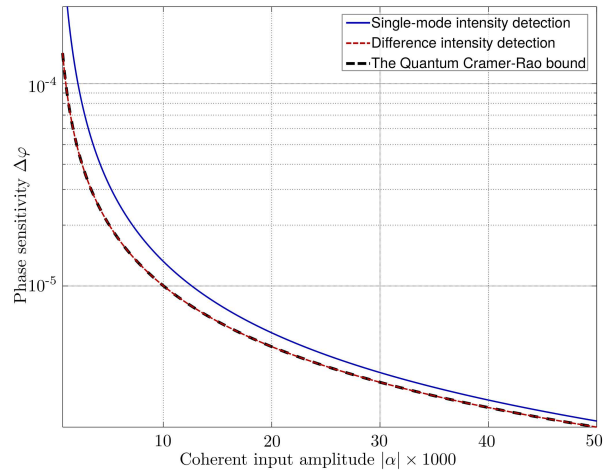}
	\caption{Phase sensitivity for the single-mode  (solid blue line), difference (dashed red line) intensity detection schemes and the quantum Cram\'er-Rao bound (thick dashed line) versus the coherent input amplitude $\vert\alpha\vert$, in the high-intensity regime. We consider a squeezing factor $r=2.3$ and we plot the optimal phases for each detection scenario. As $\vert\alpha\vert$ grows, both detection schemes approach the Cram\'er-Rao bound.}
	\label{fig:Plot_Coh_SqzVac_Diff_Single_CRB_var_alpha_high}
\end{figure}

In Fig.~\ref{fig:Plot_Coh_SqzVac_Diff_Single_CRB_var_alpha_high} we plot both both $\Delta\varphi_\mathrm{diff}$ and $\Delta\varphi_\mathrm{sing}$ in the high $\vert\alpha\vert$ regime. We conclude that if $\vert\alpha\vert^2\gg\vert\alpha\vert\gg\sinh{r}$, both detection schemes have a similar sensitivity, close to the QCRB. This agrees with the results of ref.~\cite{Gar17}.

As already mentioned, the optimum phase shift inside the Mach-Zehnder interferometer for a difference intensity detection scheme is constant, ${\varphi_\mathrm{diff}^{\mathrm{opt}}=\pi/2+k\pi}$. In this case each output port receives roughly half of the (large) input power -- this regime is clearly not desirable for the detectors.

% ---------------------------------------------------------
% ------------------------ FIGURE --- FIGURE --- FIGURE ---
% ---------------------------------------------------------

For a single-mode intensity detection scheme $\varphi_\mathrm{sing}^{\mathrm{opt}}$ is given by eq.~\eqref{eq:phi_optimal_CSV_single_DET}. Moreover, in this scenario the port 4 is almost ``dark'', and consequently we can use extremely sensitive PIN photodiodes. Almost all power exits through the output $5$ and can be discarded or used for a feedback loop to stabilise the input laser. This is the crucial difference between the two schemes in the high intensity regime, similarly to the single and double coherent input cases (see Sections \ref{sec:coherent_input} and \ref{sec:dual_coh_input}).

In this paper we did not consider losses or decoherence. The impact of losses on various scenarios has been discussed extensively in the literature \cite{Dor09,Ono10,Gar17,Dem12}. In the following we briefly discuss their effect in the high-intensity regime. Following \cite{Ono10}, in the case of a coherent input we can replace $\alpha \to \alpha \sqrt{1-\sigma}$ resulting in a quantum Cram\'er-Rao bound $\Delta\varphi_{QCRB}=1/\vert\alpha\vert \sqrt{1-\sigma}
$. The effect of small losses ($\sigma\ll1$) is marginal for a coherent source. In the case of coherent $\otimes$ squeezed vacuum input we have \cite{Ono10} the Cram\'er-Rao bound
\begin{equation}
\label{eq:Delta_QCRB_losses}
\Delta\varphi_{QCRB}^{\textrm{loss}}
\approx\frac{\sqrt{\sigma+(1-\sigma)e^{-2r}}}{\sqrt{\left(1-\sigma\right)\vert\alpha\vert^2+\sigma\left(1-\sigma\right)\sinh^2r}}
\end{equation}
The effect of losses is obvious for high squeezing factors, the numerator of equation \eqref{eq:Delta_QCRB_losses} being reduced to $\sqrt{\sigma}$, thus losing the exponential factor brought from the squeezing of the input vacuum. Nonetheless, for the high intensity regime discussed in this paper we have $\vert\alpha\vert/\sinh{r}\gg1$ and the effect of small losses is rather limited because $\sigma\ll(1-\sigma)e^{-2r}$.

For simplicity, we did not consider the $1/\sqrt{m}$ scaling for all phase sensitivities throughout this paper, where $m$ is the number of repeated experiments.

We summarized our results in Table \ref{tab:Sensitivities}.

% ---------------------------------------------------------
% ---------------- C O N C L U S I O N S ------------------
% ---------------------------------------------------------
\section{Conclusions}
\label{sec:conclusions}

The sensitivity of a Mach-Zehnder interferometer depends on both the input state and the detection setup. To achieve the best sensitivity we need to find the optimum working point(s) of the interferometer. 

For single coherent and double-coherent input, both detection setups achieve the QCRB, although at different values of $\varphi$. The double-coherent input allows us to experimentally tune the point of maximum sensitivity by adjusting the relative intensity of the two coherent states. This is an advantage over other methods involving mechanically adjusted setups.

In the high intensity regime all three input states (coherent, double coherent and coherent plus squeezed vacuum) give similar phase sensitivity, at or close to the QCRB. The optimum working point for the single-mode intensity detection has an almost ``dark'' output port. This ensures one can use highly-efficient PIN photodiodes and thus avoid potential problems of over-heating or blinding the photo-detectors. We expect that our results will lead to more sensitive detection systems for interferometry in the high-power regime.

\begin{acknowledgments}

S.A. acknowledges that this work has been supported by the Extreme Light Infrastructure Nuclear Physics (ELI-NP) Phase II, a project co-financed by the Romanian Government and the European Union through the European Regional Development Fund and the Competitiveness Operational Programme (1/07.07.2016, COP, ID 1334). R.I.~acknowledges support from a grant of the Romanian Ministry of Research and Innovation, PCCDI-UEFISCDI, project number PN-III-P1-1.2-PCCDI-2017-0338/79PCCDI/2018, within PNCDI III and PN 18090101/2018.

\end{acknowledgments}

% #########################################################
% ################        APPENDIX         ################
% #########################################################

\appendix

% ---------------------------------------------------------
% ----- DOUBLE COHERENT INPUT --- QCRB COMPUTATION --------
% ---------------------------------------------------------
\section{Quantum Cram\'er-Rao bound for a double-coherent input}
\label{sec:app:Dual_coh_QCRB}
Following reference \cite{Lan13} we consider the general case where each arm of the MZI contains a phase-shift ($\varphi_1$ and, respectively, $\varphi_2$). The estimation is treated as a general two parameter problem. We define the $2\times2$ Fisher information matrix:
\begin{equation}
\label{eq:app: Fisher_matrix}
   \mathcal{F}=
  \left[ {\begin{array}{cc}
  \mathcal{F}_{++} & \mathcal{F}_{+-} \\
   \mathcal{F}_{-+} & \mathcal{F}_{--} \\
  \end{array} } \right]
\end{equation}
where 
\begin{equation}
\label{eq:app:Fisher_matrix_elements}
\mathcal{F}_{ij}=4Re(\langle\partial_i\psi\vert\partial_j\psi\rangle-\langle\partial_i\psi\vert\psi\rangle
 \langle\psi\vert\partial_j\psi\rangle)
\end{equation} 
with $i, j= \pm$ and ${\varphi_\pm= \varphi_1\pm\varphi_2}$. From this matrix we can easily compute the QCRB:
\begin{equation}
\langle\Delta\varphi_i\Delta\varphi_j\rangle\geq(\mathcal{F}^{-1})_{ij}
\end{equation}
The state $\vert\psi\rangle$ in equation \eqref{eq:app:Fisher_matrix_elements} is:
\begin{equation}
\vert\psi\rangle=e^{-i\frac{\varphi_+}{2}(a_2^\dagger a_2+a_3^\dagger a_3)}e^{-i\frac{\varphi_-}{2}(a_2^\dagger a_2-a_3^\dagger a_3)}\vert\psi_{23}\rangle
\end{equation}
where $\vert\psi_{23}\rangle= U_{BS} \vert\alpha_1\beta_0\rangle$ is the state after the first beam splitter and $U_{BS}= e^{-i\pi/4(\hat{a}_0^\dagger\hat{a}_0+\hat{a}_1^\dagger\hat{a}_1)}$ is the unitary transformation of $BS_1$.

The elements of $\mathcal F$ are:
$\mathcal{F}_{++}= \vert\alpha\vert^2+\vert\beta\vert^2$,
${\mathcal{F}_{+-}=\mathcal{F}_{-+}=-2\vert\alpha\beta\vert\sin \Delta\theta}$ and $\mathcal{F}_{--}= \vert\alpha\vert^2+\vert\beta\vert^2$.

We are interested in the phase difference between the two arms, i.e., $\langle(\Delta\varphi_-)^2\rangle\geq(\mathcal{F}^{-1})_{--}$, for which we obtain the following QCRB:
\begin{equation}
\Delta\varphi_{QCRB}\geq
\frac{1}{\sqrt{\vert\alpha\vert^2+\vert\beta\vert^2-\frac{4\vert\alpha\beta\vert^2\sin^2 \Delta\theta}{\vert\alpha\vert^2+\vert\beta\vert^2}}}
\end{equation}
which is equivalent to eq.~\eqref{eq:Delta_varphi_QCRB_Dual_coh} with $\varpi= |\beta|/|\alpha|$.

% ---------------------------------------------------------
% --- DIFFERENCE DETECTION ---- VARIANCE COMPUTATION ------
% ---------------------------------------------------------

% ----- BEGIN    W I D E T E X T ----- BEGIN WIDETEXT -----
\begin{widetext}
\section{Calculation of the output variance}
\label{sec:app:variance_calculation}
Here we compute the averages $\langle \hat{N}^2 \rangle$ needed in the paper. For a difference intensity detection scheme, from eqs.~\eqref{eq:N_d_operator_DEFINITION} and \eqref{eq:field_op_transf_MZI} we obtain the expression of $\hat{N}_d^2$ as a function of input operators $a^\dag_0, a^\dag_1$. After a long but straightforward calculation we obtain the final, normally ordered expression
%\begin{widetext}

\begin{eqnarray}
\label{eq:N_d_SQUARED_FINAL}
% ---- SIMPLIFY 2
\langle\hat{N}_d^2\rangle
=\cos^2\varphi\langle\hat{a}_0^\dagger\hat{a}_0^\dagger\hat{a}_0\hat{a}_0\rangle
-2\cos\left(2\varphi\right)\langle\hat{a}_0^\dagger\hat{a}_0\hat{a}_1^\dagger\hat{a}_1\rangle
% ------ comment below
%\qquad\quad
%\nonumber\\
+\cos^2\varphi\langle\hat{a}_1^\dagger\hat{a}_1^\dagger\hat{a}_1\hat{a}_1\rangle
+\langle\hat{a}_0^\dagger\hat{a}_0\rangle
+\langle\hat{a}_1^\dagger\hat{a}_1\rangle
+\sin^2\varphi\langle\hat{a}_0^2(\hat{a}_1^\dagger)^2\rangle
% ------
\nonumber\\
+\sin^2\varphi\langle(\hat{a}_0^\dagger)^2\hat{a}_1^2\rangle
+\sin2\varphi\langle\hat{a}_0^\dagger\hat{a}_0^2\hat{a}_1^\dagger\rangle
+\sin2\varphi\langle(\hat{a}_0^\dagger)^2\hat{a}_0\hat{a}_1\rangle
% ------------- comment below
%\nonumber\\
-\sin2\varphi\langle\hat{a}_0(\hat{a}_1^\dagger)^2\hat{a}_1\rangle
-\sin2\varphi\langle\hat{a}_0^\dagger\hat{a}_1^\dagger\hat{a}_1^2\rangle
\qquad
\end{eqnarray}
%\end{widetext}

% ---------------------------------------------------------
% ------ SINGLE DETECTOR ---- VARIANCE COMPUTATION --------
% ---------------------------------------------------------

For the single-mode intensity detection setup, the calculation of $\langle\hat{N}_4^2\rangle$ is similar and we obtain

\begin{eqnarray}
\label{eq:N_4_SQUARED_FINAL}
% --------------------------------------
\langle\hat{N}_4^2\rangle
=\sin^4\left(\frac{\varphi}{2}\right)\langle\hat{a}_0^\dagger\hat{a}_0^\dagger\hat{a}_0\hat{a}_0\rangle
+\cos^4\left(\frac{\varphi}{2}\right)\langle\hat{a}_1^\dagger\hat{a}_1^\dagger\hat{a}_1\hat{a}_1\rangle
+\sin^2\varphi\langle\hat{a}_0^\dagger\hat{a}_0\hat{a}_1^\dagger\hat{a}_1\rangle
% -----%-----
+\sin^2\left(\frac{\varphi}{2}\right)\langle\hat{a}_0^\dagger\hat{a}_0\rangle
+\cos^2\left(\frac{\varphi}{2}\right)\langle\hat{a}_1^\dagger\hat{a}_1\rangle
\nonumber\\
%
%----- SQUARE terms
+\frac{\sin^2\varphi}{4}\langle\hat{a}_0^2(\hat{a}_1^\dagger)^2\rangle
+\frac{\sin^2\varphi}{4}\langle(\hat{a}_0^\dagger)^2\hat{a}_1^2\rangle
-\sin^2\left(\frac{\varphi}{2}\right)\sin\varphi\langle\hat{a}_0^\dagger\hat{a}_0^2\hat{a}_1^\dagger\rangle
-\sin^2\left(\frac{\varphi}{2}\right)\sin\varphi\langle(\hat{a}_0^\dagger)^2\hat{a}_0\hat{a}_1\rangle
\nonumber\\ 
% ----
-\cos^2\left(\frac{\varphi}{2}\right)\sin\varphi\langle\hat{a}_0(\hat{a}_1^\dagger)^2\hat{a}_1\rangle
-\cos^2\left(\frac{\varphi}{2}\right)\sin\varphi\langle\hat{a}_0^\dagger\hat{a}_1^\dagger\hat{a}_1^2\rangle
-\frac{\sin\varphi}{2}\langle\hat{a}_0\hat{a}_1^\dagger\rangle
-\frac{\sin\varphi}{2}\langle\hat{a}_0^\dagger\hat{a}_1\rangle\qquad
\end{eqnarray}

% --------- END    W I D E T E X T ----- END WIDETEXT -----
\end{widetext}

% ---------------------------------------------------------
% --------------- TABLE TABLE TABLE -----------------------
% ---------------------------------------------------------
\onecolumngrid

\newcommand\T{\rule{0pt}{3.6ex}}       % Top strut
\newcommand\B{\rule[-2.2ex]{0pt}{0pt}} % Bottom strut

\begin{table}%[b]
\centering
\renewcommand{\arraystretch}{1.5}
\begin{tabular}{|c|c|c|c|c|c|}
\hline
{\color{darksienna}Input} & {\color{darkblue}Quantum} &  \multicolumn{2}{c|}{Difference intensity detection} & \multicolumn{2}{c|}{Single-mode intensity detection}\\
% ----------------
\cline{3-6}
% -----------
{\color{darksienna} state} & {\color{darkblue}Cram\'er-Rao} & Optimum & Best achievable & Optimum & Best achievable\\
 % -------------------
  & {\color{darkblue}bound} & phase shift & phase sensitivity & phase shift & phase sensitivity\\
%\hline
% -----------
$\vert\psi_{in}\rangle$ & $\Delta\varphi_{QCRB}$ & $\varphi^{\mathrm{opt}}_\mathrm{diff}$ & $\Delta\varphi_\mathrm{diff}\left(\varphi_\mathrm{diff}^{\mathrm{opt}}\right)$ & $\varphi^{\mathrm{opt}}_\mathrm{sing}$ & $\Delta\varphi_\mathrm{sing} \left(\varphi_\mathrm{sing}^{\mathrm{opt}}\right)$\\
\hline
% ---------------- COHERENT
$\vert\alpha_1\rangle$ & $\frac{1}{\vert\alpha\vert}$ & $\frac{\pi}{2}$ & $\frac{1}{\vert\alpha\vert}$ & $\pi$ & $\frac{1}{\vert\alpha\vert}$ \T\B \\
\hline
% ---------------- DOUBLE COHERENT
$\vert\alpha_1\beta_0\rangle$ & $\frac{\sqrt{1+\varpi^2}}{\vert\alpha\vert\sqrt{\left(1+\varpi^2\right)^2-2\varpi\sin^2\Delta\theta}}$ &  $\pm\arctan\left(\frac{1-\varpi^2}{2\varpi}\right)$ & $\Delta\varphi_{QCRB}$ &  $\pm2\arctan\left(\frac{1}{\varpi}\right)$ & $\Delta\varphi_{QCRB}$ \T\B\\
\hline
% ---------------- COHERENT SQUEEZED VACUUM
$\vert\alpha_1\xi_0\rangle$ & $\displaystyle\frac{1}{\sqrt{\vert\alpha\vert^2e^{2r}+\sinh^2r}}$ & $\frac{\pi}{2}$ & $\frac{\sqrt{\sinh^2r+\alpha^2e^{-2r}
}}{\vert\alpha^2-\sinh^2r\vert}$ & $\pm2\arctan\left(\sqrt{\frac{\sqrt{2}\vert\alpha\vert}{\sinh{r}}}\right)$ & $\frac{\sqrt{\sinh^2r+\sqrt{2}\alpha\sinh2r+\alpha^2e^{-2r}}}
{\big\vert\alpha^2-\sinh^2r\big\vert}$ \T\B\\
\hline
\end{tabular}
\caption{\label{tab:Sensitivities}The phase sensitivity of an MZI for the input states discussed in the paper. The optimum phase shift has a period of $\pi$ (for the difference intensity detection), and respectively, $2\pi$ (for the single-mode intensity detection).}
\end{table}

\twocolumngrid

% #########################################################
% #############    B I B L I O G R A P H Y    #############
% #########################################################
%
% BibTeX users please use
% \bibliographystyle{}
% \bibliography{}
%
% APS style
\bibliographystyle{apsrev4-1}

% now include the BIB file
\bibliography{MZI_phase_sensitivity_bibtex}

\begin{thebibliography}{44}%
\makeatletter
\providecommand \@ifxundefined [1]{%
 \@ifx{#1\undefined}
}%
\providecommand \@ifnum [1]{%
 \ifnum #1\expandafter \@firstoftwo
 \else \expandafter \@secondoftwo
 \fi
}%
\providecommand \@ifx [1]{%
 \ifx #1\expandafter \@firstoftwo
 \else \expandafter \@secondoftwo
 \fi
}%
\providecommand \natexlab [1]{#1}%
\providecommand \enquote  [1]{``#1''}%
\providecommand \bibnamefont  [1]{#1}%
\providecommand \bibfnamefont [1]{#1}%
\providecommand \citenamefont [1]{#1}%
\providecommand \href@noop [0]{\@secondoftwo}%
\providecommand \href [0]{\begingroup \@sanitize@url \@href}%
\providecommand \@href[1]{\@@startlink{#1}\@@href}%
\providecommand \@@href[1]{\endgroup#1\@@endlink}%
\providecommand \@sanitize@url [0]{\catcode `\\12\catcode `\$12\catcode
  `\&12\catcode `\#12\catcode `\^12\catcode `\_12\catcode `\%12\relax}%
\providecommand \@@startlink[1]{}%
\providecommand \@@endlink[0]{}%
\providecommand \url  [0]{\begingroup\@sanitize@url \@url }%
\providecommand \@url [1]{\endgroup\@href {#1}{\urlprefix }}%
\providecommand \urlprefix  [0]{URL }%
\providecommand \Eprint [0]{\href }%
\providecommand \doibase [0]{http://dx.doi.org/}%
\providecommand \selectlanguage [0]{\@gobble}%
\providecommand \bibinfo  [0]{\@secondoftwo}%
\providecommand \bibfield  [0]{\@secondoftwo}%
\providecommand \translation [1]{[#1]}%
\providecommand \BibitemOpen [0]{}%
\providecommand \bibitemStop [0]{}%
\providecommand \bibitemNoStop [0]{.\EOS\space}%
\providecommand \EOS [0]{\spacefactor3000\relax}%
\providecommand \BibitemShut  [1]{\csname bibitem#1\endcsname}%
\let\auto@bib@innerbib\@empty
%</preamble>
\bibitem [{\citenamefont {Barnett}\ \emph {et~al.}(2003)\citenamefont
  {Barnett}, \citenamefont {Fabre},\ and\ \citenamefont {Ma{\i}tre}}]{Bar03}%
  \BibitemOpen
  \bibfield  {author} {\bibinfo {author} {\bibfnamefont {S.}~\bibnamefont
  {Barnett}}, \bibinfo {author} {\bibfnamefont {C.}~\bibnamefont {Fabre}}, \
  and\ \bibinfo {author} {\bibfnamefont {A.}~\bibnamefont {Ma{\i}tre}},\ }\href
  {\doibase 10.1140/epjd/e2003-00003-3} {\bibfield  {journal} {\bibinfo
  {journal} {Eur. Phys. J. D}\ }\textbf {\bibinfo {volume} {22}},\ \bibinfo
  {pages} {513} (\bibinfo {year} {2003})}\BibitemShut {NoStop}%
\bibitem [{\citenamefont {Caves}(1981)}]{Cav81}%
  \BibitemOpen
  \bibfield  {author} {\bibinfo {author} {\bibfnamefont {C.~M.}\ \bibnamefont
  {Caves}},\ }\href {\doibase 10.1103/PhysRevD.23.1693} {\bibfield  {journal}
  {\bibinfo  {journal} {Phys. Rev. D}\ }\textbf {\bibinfo {volume} {23}},\
  \bibinfo {pages} {1693} (\bibinfo {year} {1981})}\BibitemShut {NoStop}%
\bibitem [{\citenamefont {{The LIGO Scientific Collaboration}}(2011)}]{Aba11}%
  \BibitemOpen
  \bibfield  {author} {\bibinfo {author} {\bibnamefont {{The LIGO Scientific
  Collaboration}}},\ }\href {\doibase 10.1038/NPHYS2083} {\bibfield  {journal}
  {\bibinfo  {journal} {Nature Physics}\ }\textbf {\bibinfo {volume} {7}},\
  \bibinfo {pages} {962} (\bibinfo {year} {2011})}\BibitemShut {NoStop}%
\bibitem [{\citenamefont {Demkowicz-Dobrza\'{n}ski}\ \emph
  {et~al.}(2015)\citenamefont {Demkowicz-Dobrza\'{n}ski}, \citenamefont
  {Jarzyna},\ and\ \citenamefont {Ko\l{}ody\'{n}ski}}]{Dem15}%
  \BibitemOpen
  \bibfield  {author} {\bibinfo {author} {\bibfnamefont {R.}~\bibnamefont
  {Demkowicz-Dobrza\'{n}ski}}, \bibinfo {author} {\bibfnamefont
  {M.}~\bibnamefont {Jarzyna}}, \ and\ \bibinfo {author} {\bibfnamefont
  {J.}~\bibnamefont {Ko\l{}ody\'{n}ski}},\ }\href {\doibase
  10.1016/bs.po.2015.02.003} {\bibfield  {journal} {\bibinfo  {journal}
  {Progress in Optics}\ }\textbf {\bibinfo {volume} {60}},\ \bibinfo {pages}
  {345 } (\bibinfo {year} {2015})}\BibitemShut {NoStop}%
\bibitem [{\citenamefont {Gao}\ and\ \citenamefont {Lee}(2014)}]{Gao14}%
  \BibitemOpen
  \bibfield  {author} {\bibinfo {author} {\bibfnamefont {Y.}~\bibnamefont
  {Gao}}\ and\ \bibinfo {author} {\bibfnamefont {H.}~\bibnamefont {Lee}},\
  }\href {\doibase 10.1140/epjd/e2014-50560-1} {\bibfield  {journal} {\bibinfo
  {journal} {Eur. Phys. J. D}\ }\textbf {\bibinfo {volume} {68}},\ \bibinfo
  {pages} {347} (\bibinfo {year} {2014})}\BibitemShut {NoStop}%
\bibitem [{\citenamefont {Gerry}\ and\ \citenamefont
  {Knight}(2005)}]{GerryKnight}%
  \BibitemOpen
  \bibfield  {author} {\bibinfo {author} {\bibfnamefont {C.}~\bibnamefont
  {Gerry}}\ and\ \bibinfo {author} {\bibfnamefont {P.}~\bibnamefont {Knight}},\
  }\href {\doibase 10.1017/CBO9780511791239} {\emph {\bibinfo {title}
  {Introductory Quantum Optics}}}\ (\bibinfo  {publisher} {Cambridge University
  Press},\ \bibinfo {year} {2005})\BibitemShut {NoStop}%
\bibitem [{\citenamefont {Mandel}\ and\ \citenamefont
  {Wolf}(1995)}]{MandelWolf}%
  \BibitemOpen
  \bibfield  {author} {\bibinfo {author} {\bibfnamefont {L.}~\bibnamefont
  {Mandel}}\ and\ \bibinfo {author} {\bibfnamefont {E.}~\bibnamefont {Wolf}},\
  }\href {\doibase 10.1017/CBO9781139644105} {\emph {\bibinfo {title} {Optical
  Coherence and Quantum Optics}}}\ (\bibinfo  {publisher} {Cambridge University
  Press},\ \bibinfo {year} {1995})\BibitemShut {NoStop}%
\bibitem [{\citenamefont {Yuen}(1976)}]{Yue76}%
  \BibitemOpen
  \bibfield  {author} {\bibinfo {author} {\bibfnamefont {H.~P.}\ \bibnamefont
  {Yuen}},\ }\href {\doibase 10.1103/PhysRevA.13.2226} {\bibfield  {journal}
  {\bibinfo  {journal} {Phys. Rev. A}\ }\textbf {\bibinfo {volume} {13}},\
  \bibinfo {pages} {2226} (\bibinfo {year} {1976})}\BibitemShut {NoStop}%
\bibitem [{\citenamefont {Yurke}(1985)}]{Yur85}%
  \BibitemOpen
  \bibfield  {author} {\bibinfo {author} {\bibfnamefont {B.}~\bibnamefont
  {Yurke}},\ }\href {\doibase 10.1103/PhysRevA.32.300} {\bibfield  {journal}
  {\bibinfo  {journal} {Phys. Rev. A}\ }\textbf {\bibinfo {volume} {32}},\
  \bibinfo {pages} {300} (\bibinfo {year} {1985})}\BibitemShut {NoStop}%
\bibitem [{\citenamefont {Agarwal}(2012)}]{Aga12}%
  \BibitemOpen
  \bibfield  {author} {\bibinfo {author} {\bibfnamefont {G.~S.}\ \bibnamefont
  {Agarwal}},\ }\href {\doibase 10.1017/CBO9781139035170} {\emph {\bibinfo
  {title} {Quantum Optics}}}\ (\bibinfo  {publisher} {Cambridge University
  Press},\ \bibinfo {year} {2012})\BibitemShut {NoStop}%
\bibitem [{\citenamefont {Holland}\ and\ \citenamefont
  {Burnett}(1993)}]{Hol93}%
  \BibitemOpen
  \bibfield  {author} {\bibinfo {author} {\bibfnamefont {M.~J.}\ \bibnamefont
  {Holland}}\ and\ \bibinfo {author} {\bibfnamefont {K.}~\bibnamefont
  {Burnett}},\ }\href {\doibase 10.1103/PhysRevLett.71.1355} {\bibfield
  {journal} {\bibinfo  {journal} {Phys. Rev. Lett.}\ }\textbf {\bibinfo
  {volume} {71}},\ \bibinfo {pages} {1355} (\bibinfo {year}
  {1993})}\BibitemShut {NoStop}%
\bibitem [{\citenamefont {Paris}(1995)}]{Par95}%
  \BibitemOpen
  \bibfield  {author} {\bibinfo {author} {\bibfnamefont {M.~G.}\ \bibnamefont
  {Paris}},\ }\href {\doibase 10.1016/0375-9601(95)00235-U} {\bibfield
  {journal} {\bibinfo  {journal} {Physics Letters A}\ }\textbf {\bibinfo
  {volume} {201}},\ \bibinfo {pages} {132 } (\bibinfo {year}
  {1995})}\BibitemShut {NoStop}%
\bibitem [{\citenamefont {{The LIGO Scientific Collaboration}}(2013)}]{LIGO13}%
  \BibitemOpen
  \bibfield  {author} {\bibinfo {author} {\bibnamefont {{The LIGO Scientific
  Collaboration}}},\ }\href {\doibase 10.1038/NPHOTON.2013.177} {\bibfield
  {journal} {\bibinfo  {journal} {Nature Photonics}\ }\textbf {\bibinfo
  {volume} {7}},\ \bibinfo {pages} {616} (\bibinfo {year} {2013})}\BibitemShut
  {NoStop}%
\bibitem [{\citenamefont {Xiao}\ \emph {et~al.}(1987)\citenamefont {Xiao},
  \citenamefont {Wu},\ and\ \citenamefont {Kimble}}]{Xia87}%
  \BibitemOpen
  \bibfield  {author} {\bibinfo {author} {\bibfnamefont {M.}~\bibnamefont
  {Xiao}}, \bibinfo {author} {\bibfnamefont {L.-A.}\ \bibnamefont {Wu}}, \ and\
  \bibinfo {author} {\bibfnamefont {H.~J.}\ \bibnamefont {Kimble}},\ }\href
  {\doibase 10.1103/PhysRevLett.59.278} {\bibfield  {journal} {\bibinfo
  {journal} {Phys. Rev. Lett.}\ }\textbf {\bibinfo {volume} {59}},\ \bibinfo
  {pages} {278} (\bibinfo {year} {1987})}\BibitemShut {NoStop}%
\bibitem [{\citenamefont {Breitenbach}\ \emph {et~al.}(1998)\citenamefont
  {Breitenbach}, \citenamefont {Illuminati}, \citenamefont {Schiller},\ and\
  \citenamefont {Mlynek}}]{Bre98}%
  \BibitemOpen
  \bibfield  {author} {\bibinfo {author} {\bibfnamefont {G.}~\bibnamefont
  {Breitenbach}}, \bibinfo {author} {\bibfnamefont {F.}~\bibnamefont
  {Illuminati}}, \bibinfo {author} {\bibfnamefont {S.}~\bibnamefont
  {Schiller}}, \ and\ \bibinfo {author} {\bibfnamefont {J.}~\bibnamefont
  {Mlynek}},\ }\href {\doibase 10.1209/epl/i1998-00456-2} {\bibfield  {journal}
  {\bibinfo  {journal} {EPL (Europhysics Letters)}\ }\textbf {\bibinfo {volume}
  {44}},\ \bibinfo {pages} {192} (\bibinfo {year} {1998})}\BibitemShut
  {NoStop}%
\bibitem [{\citenamefont {Vahlbruch}\ \emph {et~al.}(2008)\citenamefont
  {Vahlbruch}, \citenamefont {Mehmet}, \citenamefont {Chelkowski},
  \citenamefont {Hage}, \citenamefont {Franzen}, \citenamefont {Lastzka},
  \citenamefont {Go\ss{}ler}, \citenamefont {Danzmann},\ and\ \citenamefont
  {Schnabel}}]{Vah10}%
  \BibitemOpen
  \bibfield  {author} {\bibinfo {author} {\bibfnamefont {H.}~\bibnamefont
  {Vahlbruch}}, \bibinfo {author} {\bibfnamefont {M.}~\bibnamefont {Mehmet}},
  \bibinfo {author} {\bibfnamefont {S.}~\bibnamefont {Chelkowski}}, \bibinfo
  {author} {\bibfnamefont {B.}~\bibnamefont {Hage}}, \bibinfo {author}
  {\bibfnamefont {A.}~\bibnamefont {Franzen}}, \bibinfo {author} {\bibfnamefont
  {N.}~\bibnamefont {Lastzka}}, \bibinfo {author} {\bibfnamefont
  {S.}~\bibnamefont {Go\ss{}ler}}, \bibinfo {author} {\bibfnamefont
  {K.}~\bibnamefont {Danzmann}}, \ and\ \bibinfo {author} {\bibfnamefont
  {R.}~\bibnamefont {Schnabel}},\ }\href {\doibase
  10.1103/PhysRevLett.100.033602} {\bibfield  {journal} {\bibinfo  {journal}
  {Phys. Rev. Lett.}\ }\textbf {\bibinfo {volume} {100}},\ \bibinfo {pages}
  {033602} (\bibinfo {year} {2008})}\BibitemShut {NoStop}%
\bibitem [{\citenamefont {Vahlbruch}\ \emph {et~al.}(2016)\citenamefont
  {Vahlbruch}, \citenamefont {Mehmet}, \citenamefont {Danzmann},\ and\
  \citenamefont {Schnabel}}]{Vah16}%
  \BibitemOpen
  \bibfield  {author} {\bibinfo {author} {\bibfnamefont {H.}~\bibnamefont
  {Vahlbruch}}, \bibinfo {author} {\bibfnamefont {M.}~\bibnamefont {Mehmet}},
  \bibinfo {author} {\bibfnamefont {K.}~\bibnamefont {Danzmann}}, \ and\
  \bibinfo {author} {\bibfnamefont {R.}~\bibnamefont {Schnabel}},\ }\href
  {\doibase 10.1103/PhysRevLett.117.110801} {\bibfield  {journal} {\bibinfo
  {journal} {Phys. Rev. Lett.}\ }\textbf {\bibinfo {volume} {117}},\ \bibinfo
  {pages} {110801} (\bibinfo {year} {2016})}\BibitemShut {NoStop}%
\bibitem [{\citenamefont {Wakui}\ \emph {et~al.}(2014)\citenamefont {Wakui},
  \citenamefont {Eto}, \citenamefont {Benichi}, \citenamefont {Izumi},
  \citenamefont {Yanagida}, \citenamefont {Ema}, \citenamefont {Numata},
  \citenamefont {Fukuda}, \citenamefont {Takeoka},\ and\ \citenamefont
  {Sasaki}}]{Wak14}%
  \BibitemOpen
  \bibfield  {author} {\bibinfo {author} {\bibfnamefont {K.}~\bibnamefont
  {Wakui}}, \bibinfo {author} {\bibfnamefont {Y.}~\bibnamefont {Eto}}, \bibinfo
  {author} {\bibfnamefont {H.}~\bibnamefont {Benichi}}, \bibinfo {author}
  {\bibfnamefont {S.}~\bibnamefont {Izumi}}, \bibinfo {author} {\bibfnamefont
  {T.}~\bibnamefont {Yanagida}}, \bibinfo {author} {\bibfnamefont
  {K.}~\bibnamefont {Ema}}, \bibinfo {author} {\bibfnamefont {T.}~\bibnamefont
  {Numata}}, \bibinfo {author} {\bibfnamefont {D.}~\bibnamefont {Fukuda}},
  \bibinfo {author} {\bibfnamefont {M.}~\bibnamefont {Takeoka}}, \ and\
  \bibinfo {author} {\bibfnamefont {M.}~\bibnamefont {Sasaki}},\ }\href
  {\doibase 10.1038/srep04535} {\bibfield  {journal} {\bibinfo  {journal}
  {Scientific Reports}\ }\textbf {\bibinfo {volume} {4}},\ \bibinfo {pages}
  {4535} (\bibinfo {year} {2014})}\BibitemShut {NoStop}%
\bibitem [{\citenamefont {Ou}(1996)}]{Ou96}%
  \BibitemOpen
  \bibfield  {author} {\bibinfo {author} {\bibfnamefont {Z.~Y.}\ \bibnamefont
  {Ou}},\ }\href {\doibase 10.1103/PhysRevLett.77.2352} {\bibfield  {journal}
  {\bibinfo  {journal} {Phys. Rev. Lett.}\ }\textbf {\bibinfo {volume} {77}},\
  \bibinfo {pages} {2352} (\bibinfo {year} {1996})}\BibitemShut {NoStop}%
\bibitem [{\citenamefont {Giovannetti}\ \emph {et~al.}(2004)\citenamefont
  {Giovannetti}, \citenamefont {Lloyd},\ and\ \citenamefont {Maccone}}]{Gio04}%
  \BibitemOpen
  \bibfield  {author} {\bibinfo {author} {\bibfnamefont {V.}~\bibnamefont
  {Giovannetti}}, \bibinfo {author} {\bibfnamefont {S.}~\bibnamefont {Lloyd}},
  \ and\ \bibinfo {author} {\bibfnamefont {L.}~\bibnamefont {Maccone}},\ }\href
  {\doibase 10.1126/science.1104149} {\bibfield  {journal} {\bibinfo  {journal}
  {Science}\ }\textbf {\bibinfo {volume} {306}},\ \bibinfo {pages} {1330}
  (\bibinfo {year} {2004})}\BibitemShut {NoStop}%
\bibitem [{\citenamefont {Pezz\'e}\ and\ \citenamefont {Smerzi}(2009)}]{Pez09}%
  \BibitemOpen
  \bibfield  {author} {\bibinfo {author} {\bibfnamefont {L.}~\bibnamefont
  {Pezz\'e}}\ and\ \bibinfo {author} {\bibfnamefont {A.}~\bibnamefont
  {Smerzi}},\ }\href {\doibase 10.1103/PhysRevLett.102.100401} {\bibfield
  {journal} {\bibinfo  {journal} {Phys. Rev. Lett.}\ }\textbf {\bibinfo
  {volume} {102}},\ \bibinfo {pages} {100401} (\bibinfo {year}
  {2009})}\BibitemShut {NoStop}%
\bibitem [{\citenamefont {Giovannetti}\ and\ \citenamefont
  {Maccone}(2012)}]{Gio12}%
  \BibitemOpen
  \bibfield  {author} {\bibinfo {author} {\bibfnamefont {V.}~\bibnamefont
  {Giovannetti}}\ and\ \bibinfo {author} {\bibfnamefont {L.}~\bibnamefont
  {Maccone}},\ }\href {\doibase 10.1103/PhysRevLett.108.210404} {\bibfield
  {journal} {\bibinfo  {journal} {Phys. Rev. Lett.}\ }\textbf {\bibinfo
  {volume} {108}},\ \bibinfo {pages} {210404} (\bibinfo {year}
  {2012})}\BibitemShut {NoStop}%
\bibitem [{\citenamefont {Pezz\'e}\ and\ \citenamefont {Smerzi}(2008)}]{Pez08}%
  \BibitemOpen
  \bibfield  {author} {\bibinfo {author} {\bibfnamefont {L.}~\bibnamefont
  {Pezz\'e}}\ and\ \bibinfo {author} {\bibfnamefont {A.}~\bibnamefont
  {Smerzi}},\ }\href {\doibase 10.1103/PhysRevLett.100.073601} {\bibfield
  {journal} {\bibinfo  {journal} {Phys. Rev. Lett.}\ }\textbf {\bibinfo
  {volume} {100}},\ \bibinfo {pages} {073601} (\bibinfo {year}
  {2008})}\BibitemShut {NoStop}%
\bibitem [{\citenamefont {Lang}\ and\ \citenamefont {Caves}(2013)}]{Lan13}%
  \BibitemOpen
  \bibfield  {author} {\bibinfo {author} {\bibfnamefont {M.~D.}\ \bibnamefont
  {Lang}}\ and\ \bibinfo {author} {\bibfnamefont {C.~M.}\ \bibnamefont
  {Caves}},\ }\href {\doibase 10.1103/PhysRevLett.111.173601} {\bibfield
  {journal} {\bibinfo  {journal} {Phys. Rev. Lett.}\ }\textbf {\bibinfo
  {volume} {111}},\ \bibinfo {pages} {173601} (\bibinfo {year}
  {2013})}\BibitemShut {NoStop}%
\bibitem [{\citenamefont {Lang}\ and\ \citenamefont {Caves}(2014)}]{Lan14}%
  \BibitemOpen
  \bibfield  {author} {\bibinfo {author} {\bibfnamefont {M.~D.}\ \bibnamefont
  {Lang}}\ and\ \bibinfo {author} {\bibfnamefont {C.~M.}\ \bibnamefont
  {Caves}},\ }\href {\doibase 10.1103/PhysRevA.90.025802} {\bibfield  {journal}
  {\bibinfo  {journal} {Phys. Rev. A}\ }\textbf {\bibinfo {volume} {90}},\
  \bibinfo {pages} {025802} (\bibinfo {year} {2014})}\BibitemShut {NoStop}%
\bibitem [{\citenamefont {Sparaciari}\ \emph {et~al.}(2015)\citenamefont
  {Sparaciari}, \citenamefont {Olivares},\ and\ \citenamefont {Paris}}]{Spa15}%
  \BibitemOpen
  \bibfield  {author} {\bibinfo {author} {\bibfnamefont {C.}~\bibnamefont
  {Sparaciari}}, \bibinfo {author} {\bibfnamefont {S.}~\bibnamefont
  {Olivares}}, \ and\ \bibinfo {author} {\bibfnamefont {M.~G.~A.}\ \bibnamefont
  {Paris}},\ }\href {\doibase 10.1364/JOSAB.32.001354} {\bibfield  {journal}
  {\bibinfo  {journal} {J. Opt. Soc. Am. B}\ }\textbf {\bibinfo {volume}
  {32}},\ \bibinfo {pages} {1354} (\bibinfo {year} {2015})}\BibitemShut
  {NoStop}%
\bibitem [{\citenamefont {Sparaciari}\ \emph {et~al.}(2016)\citenamefont
  {Sparaciari}, \citenamefont {Olivares},\ and\ \citenamefont {Paris}}]{Spa16}%
  \BibitemOpen
  \bibfield  {author} {\bibinfo {author} {\bibfnamefont {C.}~\bibnamefont
  {Sparaciari}}, \bibinfo {author} {\bibfnamefont {S.}~\bibnamefont
  {Olivares}}, \ and\ \bibinfo {author} {\bibfnamefont {M.~G.~A.}\ \bibnamefont
  {Paris}},\ }\href {\doibase 10.1103/PhysRevA.93.023810} {\bibfield  {journal}
  {\bibinfo  {journal} {Phys. Rev. A}\ }\textbf {\bibinfo {volume} {93}},\
  \bibinfo {pages} {023810} (\bibinfo {year} {2016})}\BibitemShut {NoStop}%
\bibitem [{\citenamefont {Pezz\'e}\ \emph {et~al.}(2007)\citenamefont
  {Pezz\'e}, \citenamefont {Smerzi}, \citenamefont {Khoury}, \citenamefont
  {Hodelin},\ and\ \citenamefont {Bouwmeester}}]{Pez07}%
  \BibitemOpen
  \bibfield  {author} {\bibinfo {author} {\bibfnamefont {L.}~\bibnamefont
  {Pezz\'e}}, \bibinfo {author} {\bibfnamefont {A.}~\bibnamefont {Smerzi}},
  \bibinfo {author} {\bibfnamefont {G.}~\bibnamefont {Khoury}}, \bibinfo
  {author} {\bibfnamefont {J.~F.}\ \bibnamefont {Hodelin}}, \ and\ \bibinfo
  {author} {\bibfnamefont {D.}~\bibnamefont {Bouwmeester}},\ }\href {\doibase
  10.1103/PhysRevLett.99.223602} {\bibfield  {journal} {\bibinfo  {journal}
  {Phys. Rev. Lett.}\ }\textbf {\bibinfo {volume} {99}},\ \bibinfo {pages}
  {223602} (\bibinfo {year} {2007})}\BibitemShut {NoStop}%
\bibitem [{\citenamefont {Gard}\ \emph {et~al.}(2017)\citenamefont {Gard},
  \citenamefont {You}, \citenamefont {Mishra}, \citenamefont {Singh},
  \citenamefont {Lee}, \citenamefont {Corbitt},\ and\ \citenamefont
  {Dowling}}]{Gar17}%
  \BibitemOpen
  \bibfield  {author} {\bibinfo {author} {\bibfnamefont {B.~T.}\ \bibnamefont
  {Gard}}, \bibinfo {author} {\bibfnamefont {C.}~\bibnamefont {You}}, \bibinfo
  {author} {\bibfnamefont {D.~K.}\ \bibnamefont {Mishra}}, \bibinfo {author}
  {\bibfnamefont {R.}~\bibnamefont {Singh}}, \bibinfo {author} {\bibfnamefont
  {H.}~\bibnamefont {Lee}}, \bibinfo {author} {\bibfnamefont {T.~R.}\
  \bibnamefont {Corbitt}}, \ and\ \bibinfo {author} {\bibfnamefont {J.~P.}\
  \bibnamefont {Dowling}},\ }\href {\doibase 10.1140/epjqt/s40507-017-0058-8}
  {\bibfield  {journal} {\bibinfo  {journal} {EPJ Quantum Technology}\ }\textbf
  {\bibinfo {volume} {4}},\ \bibinfo {pages} {4} (\bibinfo {year}
  {2017})}\BibitemShut {NoStop}%
\bibitem [{\citenamefont {Shin}\ \emph {et~al.}(1999)\citenamefont {Shin},
  \citenamefont {Kim}, \citenamefont {Park}, \citenamefont {Kim},\ and\
  \citenamefont {Park}}]{Shi99}%
  \BibitemOpen
  \bibfield  {author} {\bibinfo {author} {\bibfnamefont {J.-T.}\ \bibnamefont
  {Shin}}, \bibinfo {author} {\bibfnamefont {H.-N.}\ \bibnamefont {Kim}},
  \bibinfo {author} {\bibfnamefont {G.-D.}\ \bibnamefont {Park}}, \bibinfo
  {author} {\bibfnamefont {T.-S.}\ \bibnamefont {Kim}}, \ and\ \bibinfo
  {author} {\bibfnamefont {D.-Y.}\ \bibnamefont {Park}},\ }\href
  {http://www.osapublishing.org/josk/abstract.cfm?URI=josk-3-1-1} {\bibfield
  {journal} {\bibinfo  {journal} {J. Opt. Soc. Korea}\ }\textbf {\bibinfo
  {volume} {3}},\ \bibinfo {pages} {1} (\bibinfo {year} {1999})}\BibitemShut
  {NoStop}%
\bibitem [{\citenamefont {Nagata}\ \emph {et~al.}(2007)\citenamefont {Nagata},
  \citenamefont {Okamoto}, \citenamefont {O{\textquoteright}Brien},
  \citenamefont {Sasaki},\ and\ \citenamefont {Takeuchi}}]{Nag07}%
  \BibitemOpen
  \bibfield  {author} {\bibinfo {author} {\bibfnamefont {T.}~\bibnamefont
  {Nagata}}, \bibinfo {author} {\bibfnamefont {R.}~\bibnamefont {Okamoto}},
  \bibinfo {author} {\bibfnamefont {J.~L.}\ \bibnamefont
  {O{\textquoteright}Brien}}, \bibinfo {author} {\bibfnamefont
  {K.}~\bibnamefont {Sasaki}}, \ and\ \bibinfo {author} {\bibfnamefont
  {S.}~\bibnamefont {Takeuchi}},\ }\href {\doibase 10.1126/science.1138007}
  {\bibfield  {journal} {\bibinfo  {journal} {Science}\ }\textbf {\bibinfo
  {volume} {316}},\ \bibinfo {pages} {726} (\bibinfo {year}
  {2007})}\BibitemShut {NoStop}%
\bibitem [{\citenamefont {Afek}\ \emph {et~al.}(2010)\citenamefont {Afek},
  \citenamefont {Ambar},\ and\ \citenamefont {Silberberg}}]{Afe10}%
  \BibitemOpen
  \bibfield  {author} {\bibinfo {author} {\bibfnamefont {I.}~\bibnamefont
  {Afek}}, \bibinfo {author} {\bibfnamefont {O.}~\bibnamefont {Ambar}}, \ and\
  \bibinfo {author} {\bibfnamefont {Y.}~\bibnamefont {Silberberg}},\ }\href
  {\doibase 10.1126/science.1188172} {\bibfield  {journal} {\bibinfo  {journal}
  {Science}\ }\textbf {\bibinfo {volume} {328}},\ \bibinfo {pages} {879}
  (\bibinfo {year} {2010})}\BibitemShut {NoStop}%
\bibitem [{\citenamefont {Ataman}(2018)}]{Ata18b}%
  \BibitemOpen
  \bibfield  {author} {\bibinfo {author} {\bibfnamefont {S.}~\bibnamefont
  {Ataman}},\ }\href {\doibase 10.1103/PhysRevA.97.063811} {\bibfield
  {journal} {\bibinfo  {journal} {Phys. Rev. A}\ }\textbf {\bibinfo {volume}
  {97}},\ \bibinfo {pages} {063811} (\bibinfo {year} {2018})}\BibitemShut
  {NoStop}%
\bibitem [{\citenamefont {Yurke}\ \emph {et~al.}(1986)\citenamefont {Yurke},
  \citenamefont {McCall},\ and\ \citenamefont {Klauder}}]{Yur86}%
  \BibitemOpen
  \bibfield  {author} {\bibinfo {author} {\bibfnamefont {B.}~\bibnamefont
  {Yurke}}, \bibinfo {author} {\bibfnamefont {S.~L.}\ \bibnamefont {McCall}}, \
  and\ \bibinfo {author} {\bibfnamefont {J.~R.}\ \bibnamefont {Klauder}},\
  }\href {\doibase 10.1103/PhysRevA.33.4033} {\bibfield  {journal} {\bibinfo
  {journal} {Phys. Rev. A}\ }\textbf {\bibinfo {volume} {33}},\ \bibinfo
  {pages} {4033} (\bibinfo {year} {1986})}\BibitemShut {NoStop}%
\bibitem [{\citenamefont {Braunstein}\ and\ \citenamefont
  {Caves}(1994)}]{Bra94}%
  \BibitemOpen
  \bibfield  {author} {\bibinfo {author} {\bibfnamefont {S.~L.}\ \bibnamefont
  {Braunstein}}\ and\ \bibinfo {author} {\bibfnamefont {C.~M.}\ \bibnamefont
  {Caves}},\ }\href {\doibase 10.1103/PhysRevLett.72.3439} {\bibfield
  {journal} {\bibinfo  {journal} {Phys. Rev. Lett.}\ }\textbf {\bibinfo
  {volume} {72}},\ \bibinfo {pages} {3439} (\bibinfo {year}
  {1994})}\BibitemShut {NoStop}%
\bibitem [{\citenamefont {Paris}(2009)}]{Par09}%
  \BibitemOpen
  \bibfield  {author} {\bibinfo {author} {\bibfnamefont {M.~G.~A.}\
  \bibnamefont {Paris}},\ }\href {\doibase 10.1142/S0219749909004839}
  {\bibfield  {journal} {\bibinfo  {journal} {International Journal of Quantum
  Information}\ }\textbf {\bibinfo {volume} {07}},\ \bibinfo {pages} {125}
  (\bibinfo {year} {2009})}\BibitemShut {NoStop}%
\bibitem [{\citenamefont {Jarzyna}\ and\ \citenamefont
  {Demkowicz-Dobrza\ifmmode~\acute{n}\else \'{n}\fi{}ski}(2012)}]{Jar12}%
  \BibitemOpen
  \bibfield  {author} {\bibinfo {author} {\bibfnamefont {M.}~\bibnamefont
  {Jarzyna}}\ and\ \bibinfo {author} {\bibfnamefont {R.}~\bibnamefont
  {Demkowicz-Dobrza\ifmmode~\acute{n}\else \'{n}\fi{}ski}},\ }\href {\doibase
  10.1103/PhysRevA.85.011801} {\bibfield  {journal} {\bibinfo  {journal} {Phys.
  Rev. A}\ }\textbf {\bibinfo {volume} {85}},\ \bibinfo {pages} {011801}
  (\bibinfo {year} {2012})}\BibitemShut {NoStop}%
\bibitem [{\citenamefont {Takeoka}\ \emph {et~al.}(2017)\citenamefont
  {Takeoka}, \citenamefont {Seshadreesan}, \citenamefont {You}, \citenamefont
  {Izumi},\ and\ \citenamefont {Dowling}}]{Tak17}%
  \BibitemOpen
  \bibfield  {author} {\bibinfo {author} {\bibfnamefont {M.}~\bibnamefont
  {Takeoka}}, \bibinfo {author} {\bibfnamefont {K.~P.}\ \bibnamefont
  {Seshadreesan}}, \bibinfo {author} {\bibfnamefont {C.}~\bibnamefont {You}},
  \bibinfo {author} {\bibfnamefont {S.}~\bibnamefont {Izumi}}, \ and\ \bibinfo
  {author} {\bibfnamefont {J.~P.}\ \bibnamefont {Dowling}},\ }\href {\doibase
  10.1103/PhysRevA.96.052118} {\bibfield  {journal} {\bibinfo  {journal} {Phys.
  Rev. A}\ }\textbf {\bibinfo {volume} {96}},\ \bibinfo {pages} {052118}
  (\bibinfo {year} {2017})}\BibitemShut {NoStop}%
\bibitem [{\citenamefont {Brown}\ \emph {et~al.}(2010)\citenamefont {Brown},
  \citenamefont {Larason},\ and\ \citenamefont {Ohno}}]{Bro10}%
  \BibitemOpen
  \bibfield  {author} {\bibinfo {author} {\bibfnamefont {S.~W.}\ \bibnamefont
  {Brown}}, \bibinfo {author} {\bibfnamefont {T.~C.}\ \bibnamefont {Larason}},
  \ and\ \bibinfo {author} {\bibfnamefont {Y.}~\bibnamefont {Ohno}},\ }\href
  {\doibase 10.1088/0026-1394/47/1A/02002} {\bibfield  {journal} {\bibinfo
  {journal} {Metrologia}\ }\textbf {\bibinfo {volume} {47}},\ \bibinfo {pages}
  {02002} (\bibinfo {year} {2010})}\BibitemShut {NoStop}%
\bibitem [{\citenamefont {Schnabel}(2017)}]{Sch17}%
  \BibitemOpen
  \bibfield  {author} {\bibinfo {author} {\bibfnamefont {R.}~\bibnamefont
  {Schnabel}},\ }\href {\doibase 10.1016/j.physrep.2017.04.001} {\bibfield
  {journal} {\bibinfo  {journal} {Physics Reports}\ }\textbf {\bibinfo {volume}
  {684}},\ \bibinfo {pages} {1 } (\bibinfo {year} {2017})}\BibitemShut
  {NoStop}%
\bibitem [{\citenamefont {Li}\ \emph {et~al.}(2014)\citenamefont {Li},
  \citenamefont {Yuan}, \citenamefont {Ou},\ and\ \citenamefont
  {Zhang}}]{Li14}%
  \BibitemOpen
  \bibfield  {author} {\bibinfo {author} {\bibfnamefont {D.}~\bibnamefont
  {Li}}, \bibinfo {author} {\bibfnamefont {C.-H.}\ \bibnamefont {Yuan}},
  \bibinfo {author} {\bibfnamefont {Z.~Y.}\ \bibnamefont {Ou}}, \ and\ \bibinfo
  {author} {\bibfnamefont {W.}~\bibnamefont {Zhang}},\ }\href
  {http://stacks.iop.org/1367-2630/16/i=7/a=073020} {\bibfield  {journal}
  {\bibinfo  {journal} {New Journal of Physics}\ }\textbf {\bibinfo {volume}
  {16}},\ \bibinfo {pages} {073020} (\bibinfo {year} {2014})}\BibitemShut
  {NoStop}%
\bibitem [{\citenamefont {Dorner}\ \emph {et~al.}(2009)\citenamefont {Dorner},
  \citenamefont {Demkowicz-Dobrzanski}, \citenamefont {Smith}, \citenamefont
  {Lundeen}, \citenamefont {Wasilewski}, \citenamefont {Banaszek},\ and\
  \citenamefont {Walmsley}}]{Dor09}%
  \BibitemOpen
  \bibfield  {author} {\bibinfo {author} {\bibfnamefont {U.}~\bibnamefont
  {Dorner}}, \bibinfo {author} {\bibfnamefont {R.}~\bibnamefont
  {Demkowicz-Dobrzanski}}, \bibinfo {author} {\bibfnamefont {B.~J.}\
  \bibnamefont {Smith}}, \bibinfo {author} {\bibfnamefont {J.~S.}\ \bibnamefont
  {Lundeen}}, \bibinfo {author} {\bibfnamefont {W.}~\bibnamefont {Wasilewski}},
  \bibinfo {author} {\bibfnamefont {K.}~\bibnamefont {Banaszek}}, \ and\
  \bibinfo {author} {\bibfnamefont {I.~A.}\ \bibnamefont {Walmsley}},\ }\href
  {\doibase 10.1103/PhysRevLett.102.040403} {\bibfield  {journal} {\bibinfo
  {journal} {Phys. Rev. Lett.}\ }\textbf {\bibinfo {volume} {102}},\ \bibinfo
  {pages} {040403} (\bibinfo {year} {2009})}\BibitemShut {NoStop}%
\bibitem [{\citenamefont {Ono}\ and\ \citenamefont {Hofmann}(2010)}]{Ono10}%
  \BibitemOpen
  \bibfield  {author} {\bibinfo {author} {\bibfnamefont {T.}~\bibnamefont
  {Ono}}\ and\ \bibinfo {author} {\bibfnamefont {H.~F.}\ \bibnamefont
  {Hofmann}},\ }\href {\doibase 10.1103/PhysRevA.81.033819} {\bibfield
  {journal} {\bibinfo  {journal} {Phys. Rev. A}\ }\textbf {\bibinfo {volume}
  {81}},\ \bibinfo {pages} {033819} (\bibinfo {year} {2010})}\BibitemShut
  {NoStop}%
\bibitem [{\citenamefont {Demkowicz-Dobrza\'{n}ski}\ \emph
  {et~al.}(2012)\citenamefont {Demkowicz-Dobrza\'{n}ski}, \citenamefont
  {Ko\l{}ody\'{n}ski},\ and\ \citenamefont {Gu\c{t}\u{a}}}]{Dem12}%
  \BibitemOpen
  \bibfield  {author} {\bibinfo {author} {\bibfnamefont {R.}~\bibnamefont
  {Demkowicz-Dobrza\'{n}ski}}, \bibinfo {author} {\bibfnamefont
  {J.}~\bibnamefont {Ko\l{}ody\'{n}ski}}, \ and\ \bibinfo {author}
  {\bibfnamefont {M.}~\bibnamefont {Gu\c{t}\u{a}}},\ }\href {\doibase
  10.1038/ncomms2067} {\bibfield  {journal} {\bibinfo  {journal} {Nature
  Communications}\ }\textbf {\bibinfo {volume} {3}},\ \bibinfo {pages} {1063}
  (\bibinfo {year} {2012})}\BibitemShut {NoStop}%
\end{thebibliography}%

\end{document}